\documentclass[review]{elsarticle}

\usepackage{geometry}
\usepackage{lineno}

\usepackage{xcolor}
\usepackage{url}
\usepackage[hyperindex,breaklinks]{hyperref}
\hypersetup{linkcolor=red, citecolor=red, colorlinks=true}

\usepackage{graphicx}
\usepackage[capitalise]{cleveref}
\usepackage[caption = false]{subfig}
\usepackage{placeins} 

\modulolinenumbers[5]
\journal{Chaos, Solitons \& Fractals}

\begin{document}

\begin{frontmatter}

\title{Hysteresis and disorder-induced order in continuous kinetic-like opinion dynamics in complex networks}

\author[add1]{A. L. Oestereich}
\ead{andrelo@if.uff.br}
\author[add2]{M. A. Pires}
\ead{piresma@cbpf.br}

\author[add2,add3]{S. M. Duarte~Queir\'{o}s}
\ead{sdqueiro@cbpf.br}

\author[add1]{N. Crokidakis}
\ead{nuno@mail.if.uff.br}

%
\address[add1]{Instituto de F\'isica, Universidade Federal Fluminense, Niter\'oi/RJ, Brazil}
\address[add2]{Centro Brasileiro de Pesquisas F\'isicas, Rio de Janeiro/RJ, Brazil}
\address[add3]{Instituto Nacional de Ci\^encia e Tecnologia de Sistemas Complexos INCT-SC, Rio de Janeiro/RJ, Brazil}

\begin{abstract}

In this work we tackle a kinetic-like model of opinions dynamics in a
   networked population endued with a quenched plurality and polarization.
Additionally, we consider pairwise interactions that are restrictive, which is   modeled with a smooth bounded confidence. Our results show the interesting emergence of nonequilibrium hysteresis and heterogeneity-assisted ordering.
Such counterintuitive phenomena are robust to different types of network architectures such as random, small-world and scale-free.
\end{abstract}


\end{frontmatter}


\section{Introduction } 
\label{sec:intro}

It is nowadays largely accepted that the public debate has been moving towards a framework of increasing polarization in exchange of a diversity of opinions around the so-called common-sense\cite{barber2015causes,mccoy2018polarization,urlpoliticalpolarization2020,bottcher2020great}. At the same time, one has seen in the past   decades a surge in the utilization physical systems --- or quantitative models clearly inspired therefrom --- providing valuable answers to the evolution of social systems, namely the formation of a majority opinion or consensus depending on the architecture of the interactions between the individuals as well as its robustness with respect to perturbations~\cite{2018biswasLF}.

Modeling opinion dynamics is one the most challenging topics in the study of complex systems~\cite{sirbu2017opinion,sen2014sociophysics,2012galam,2009castellanoFL,2008galam,2017albiPTZ,2011xiaWX}. Recent reviews focus on recent advances in opinion formation modeling, coupling opinion exchange with other aspects of social interactions, like the roles of conviction, positive and negative interactions, the information each individual is exposed to, and the interplay between the topology of complex networks and the spreading of opinions~\cite{2013bellomoMT,2002HegselmannK,2010naldiPT}.

Opinion formation is a complex process depending on the information that we collect from peers or other external sources,   among which the mass media is certainly the most predominant.
A great diversity of models inspired from those in use in physics have been developed to take into account those ingredients allowing us to identify the mechanisms involved in the process of opinion formation and the understanding of their role. Based on discrete and continuous opinion variables, many authors studied the role of factors such as the inflow and outflow dynamics and distinct interaction rules, among others. Network topology was also explored as well as the presence of specific kinds of agents like contrarians, inflexibles and opportunists~\cite{sirbu2017opinion,sen2014sociophysics,2017albiPTZ,2013bellomoMT,dong2018survey,bottcher2020great}. 

The present paper is organized as follows: in Sec.~\ref{sec:litrev} we present a review the correlated literature, in Sec.~\ref{sec:model} we introduce our model, in Sec.~\ref{sec:results} we present the results and in Sec.~\ref{sec:remarks} we address our final remarks on this research.


\section{Literature review} \label{sec:litrev}

Counterintuitive phenomenology that challenges conventional wisdom in social dynamics has been reported since the early stages of this field~\cite{galam1990social}. More specifically, within the field of opinion dynamics, results that defy common sense have been reported regarding the monotonicity of the ordering or consensus achievement. Assuming only endogenous ingredients in the dynamics, and by considering a mean-field kinetic-like model with non-smooth bounded confidence,~\cite{deffuant2000mixing} it was possible to show that a moderate amount of conviction assists the global opinion, but too much ends up hindering the collective opinion~\cite{sen2012nonconservative}.   Afterwards, in Ref.~\cite{2018jedrzejewskiS} it was shown that the $q$-voter model~\cite{2009castellanoMP} with memory leads to the public opinion to be a non-monotonic function of the social temperature for individualistic societies. In Ref.~\cite{anteneodo2017symmetry} it was studied a mean-field continuous opinion dynamics and reported that increasing a   `social temperature' from a small value to a moderated magnitude lead to the strengthening of the collective opinion. Recently, Ref.~\cite{2019oestereichPC} investigated a three-state kinetic-like opinion dynamics in modular networks obtaining a non-monotonic dependence of the global opinion with the network modularity and with a neutrality-noise.

On the other hand, when there are external elements driving the system --- e.g., by means of a binary opinion dynamics governed by a local majority rule under a time-dependent external field~\cite{tessone2009diversity} --- it is possible to observe   the emergence of an optimal global opinion for a moderated amount of diversity in the agent's preferences in a way that to much diversity or the lack thereof weakens the collective behavior. 

Taking note of the aforementioned references, we have noticed that opinion dynamics models have considered either agents on full graphs with continuous opinions and unrestricted interactions~\cite{anteneodo2017symmetry} or non-smooth bounded confidence~\cite{sen2012nonconservative} binary opinion states~\cite{2018jedrzejewskiS}, or else 3-state opinion approaches on modular complex networks~\cite{2019oestereichPC}. Thus, models for agents with continuous-valued opinions, restrictive interactions, individual preferences with two sources of diversity on complex network architectures is absent in the literature. 

In our model, we consider that individuals have individual preference. For that, we employ a double gaussian, which allows tuning between different sources of disorder, namely polarization and (standard) plurality around some default value. As shown in Ref.~\cite{2016queiros}, such a distinction between types of diversity is very important since they lead to different effects.   More specifically, personal preferences are implemented by introducing an individual additive field to the equation of opinion formation. 
  
This setting assumes that voters evaluate relevant information in a more or less impartial way, but there is strong behavioral evidence of motivated reasoning and partisan bias. It is known (in political psychology) that personality traits correlate with political preference. Namely, it is asserted~\cite{2007jostGPO,2009jostFN,2010hirshYXP} --- within statistical significance --- that Conservatives tend to be higher in conscientiousness and politeness, whereas Liberals are higher in openness-intellect and compassion. So much so, that neuroanatomical differences were also found to correlate with political preference~\cite{2014jostNAB}. That said, as personal preferences play an important role in the public debate, they should be taken into account in its models as well.

\section{Model} \label{sec:model}

Based on a local-global kinetic-like model of opinions presented Refs.~\cite{2010lallouacheCCC,2011biswasCC}, we consider a model where the opinion, $o$, of each agent, $i$, evolves according to
\begin{equation}
      o_i (t+1) = o_i(t) + \epsilon_1 g(t) o_j(t)   + \epsilon_2 O(t) + h_i.
      \label{eq:evol}
\end{equation}
Explicitly, we consider that the opinion of an agent depends on her previous opinion $ o_i(t)$, a peer pressure $o_j(t)$ of a randomly picked neighbor $j$ which is modulated by random heterogeneity,~\cite{2010lallouacheCCC} $\epsilon_1 \in [0,1]$ and a smooth bounded confidence,~\cite{deffuant2004modelling}
\begin{equation}
  g(t) = e^{-r[ o_i(t) - o_j(t)]^2 }
      \label{eq:smoothboundedconfidence}
\end{equation}
where $r$ is a decay rate that calibrates the magnitude of the relative peer opinion. This term produces restrictive interactions since it bounds the impact of the opinion of the neighbor $j$ in the future opinion of agent $i$. The third term on the right-hand-side of \cref{eq:evol} aims at describing effective perception of agent $i$ over the global opinion that is weighed by annealed stochastic heterogeneity   $\epsilon_2 \sim U(0,1)$~\cite{2011biswasCC}.
Last, we represent the bias of the agent --- established on their general beliefs and traits --- by the variable $h$. Herein, that endogenous field is randomly selected according to a double gaussian probability distribution,
\begin{equation}
      p(h_i) = \frac{1}{\sqrt{2\pi} \, \sigma} \sum _\pm \exp \left[ - \frac{(h_i \pm \mu )^2}{\sigma ^2} \right].
      \label{eq:probDist}
\end{equation}
The limiting case $\sigma=0$ on \cref{eq:probDist} corresponds to a bimodal distribution, i.e., two delta functions centered at $h_{i}=\pm \mu$. Since such features-- individual preferences -- tend do be stable trough individuals (adult) lives we treat $\left\{h_i \right\}$ as quenched.   The probability density function (\ref{eq:probDist}) has two free parameters of the model and was selected because it allows us to study the impact of different sorts of heterogeneity, namely plurality --- quantified by $\sigma $ --- and polarization, which is gauged by the value of $\mu$~\cite{2016queiros}. The approach of considering individual features as an additive field was first championed by Galam in a simple Ising-like model adapted to survey group decision making~\cite{1997galam}.

If $r=\mu=\sigma=0$ we recover the model by Biswas et al~\cite{2011biswasCC} that has no bounded confidence and no individual's preference while keeping the term related to the partial knowledge of the global opinion.


The initial state of the system is assumed to be fully disordered(ordered); in other words, at first the opinions are drawn from the uniform distribution in the range $[-1,1]$ ($o_i = 1 \; \forall i$). Afterwards, the algorithm goes as follows: at each time step $t$ we select an individual $i$ and one of its neighbours $j$, and update the opinion $o_i (t)$ according to~\cref{eq:evol}. If opinion $o_i(t+1)$ exceeds the extreme values then it is reinserted at its corresponding limiting value, that is, $1$ for $o_i(t+1) > 1$ and $-1$ for $o_i(t+1) < -1$. $N$ of these updates define a unit of time. The structural disorder was implemented through three   paradigmatic networks: Erd\H{o}s-R\'enyi (ER)~\cite{1959erdosR}, Barab\'asi-Albert (BA)~\cite{1999barabasiA} and Watts-Strogatz (WS)~\cite{1998wattsS}.

To characterize the collective stance of the group of agents we use
\begin{equation}
      O = \frac{1}{N} \left| \sum_{i=1} ^{N} o_i \, \right|,
      \label{eq:oParam}
\end{equation}
which in critical phenomena parlance acts as the order parameter of this system so that a nontrivial collective stance is achieved when $O \neq 0$; on the other hand, the state $O = 0$ corresponds the trivial collective stance, which at its best, corresponds to the classical 'we agree to disagree' situation.
Physically, the former is a ordered state whereas the latter is a disordered one. \Cref{eq:oParam} allows addressing the link between the micro-level features given by $o_i$ and the aggregate macro-level outcome $O$, which is an issue of fundamental importance in social sciences~\cite{stadtfeld2015micro}.

\section{\label{sec:results}Results and discussion}

\begin{figure*}[t]
      \centering
      \subfloat[ER, $\sigma = 0.00$]{\includegraphics[width=0.33\textwidth]{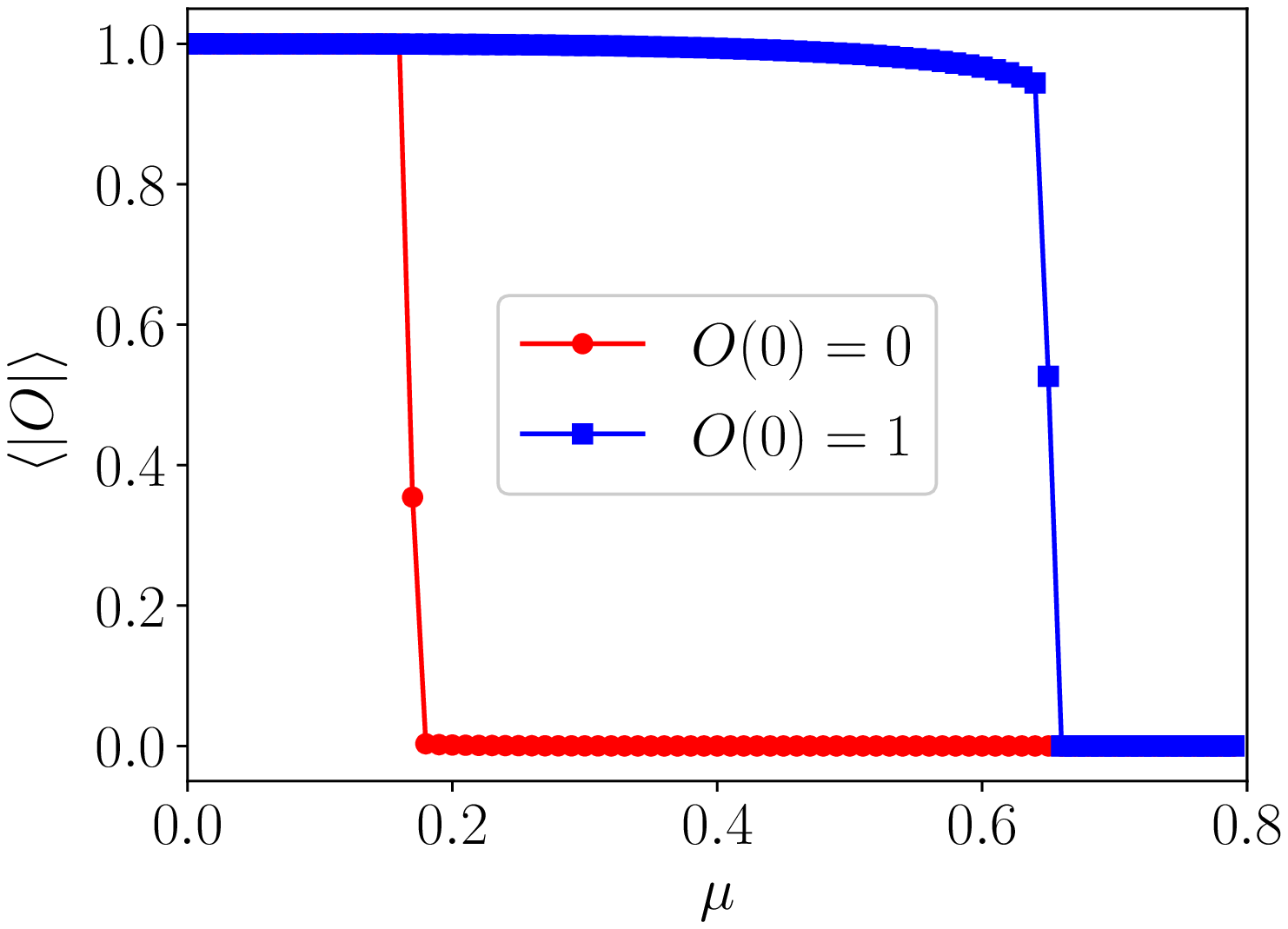}} 
      \subfloat[BA, $\sigma = 0.00$]{\includegraphics[width=0.33\textwidth]{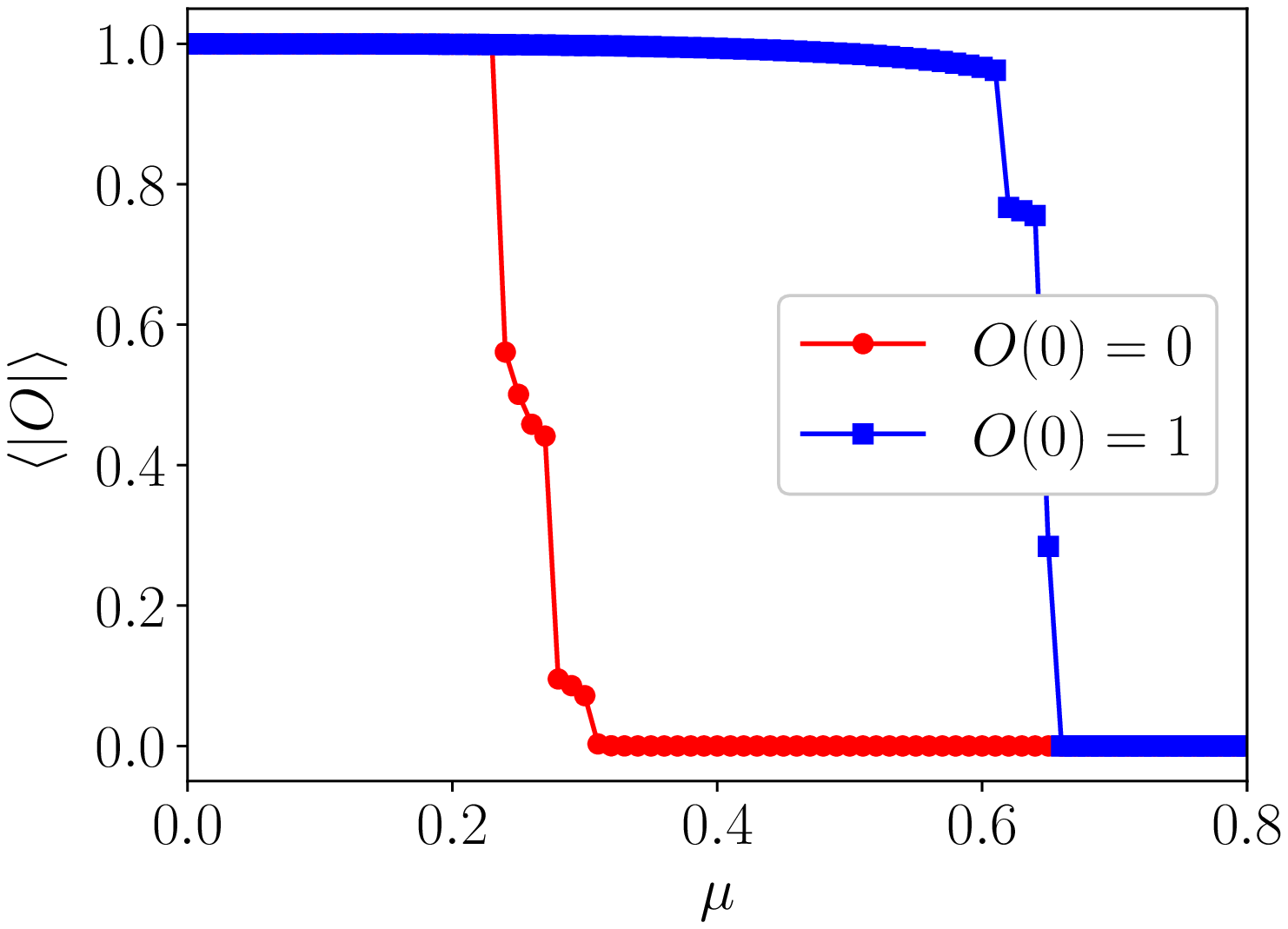}} 
      \subfloat[WS, $\sigma = 0.00$]{\includegraphics[width=0.33\textwidth]{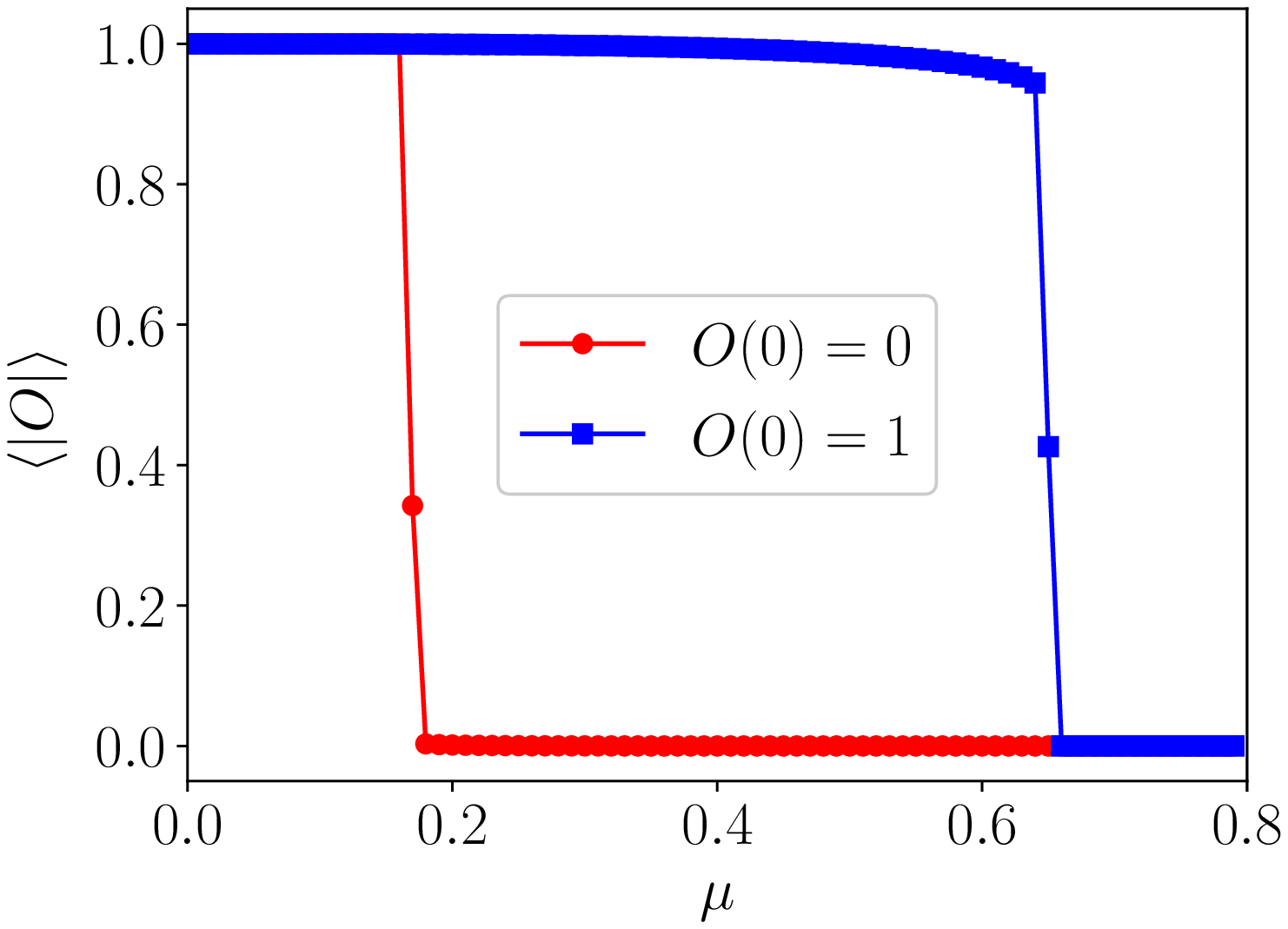}}
      \caption{Collective stance $O$ versus $\mu$ for a model of $N=10^4$ agents following on the different networks with $\langle k \rangle =50$, $\sigma = 0$ and $r = 1/4$. The results are for both initial conditions. Here we can see the hysteresis curve that indicates a discontinuous phase transition.}
      \label{fig:Oxsigma_l_1}
\medskip
      \centering
      \subfloat[ER]{\includegraphics[width=0.33\textwidth]{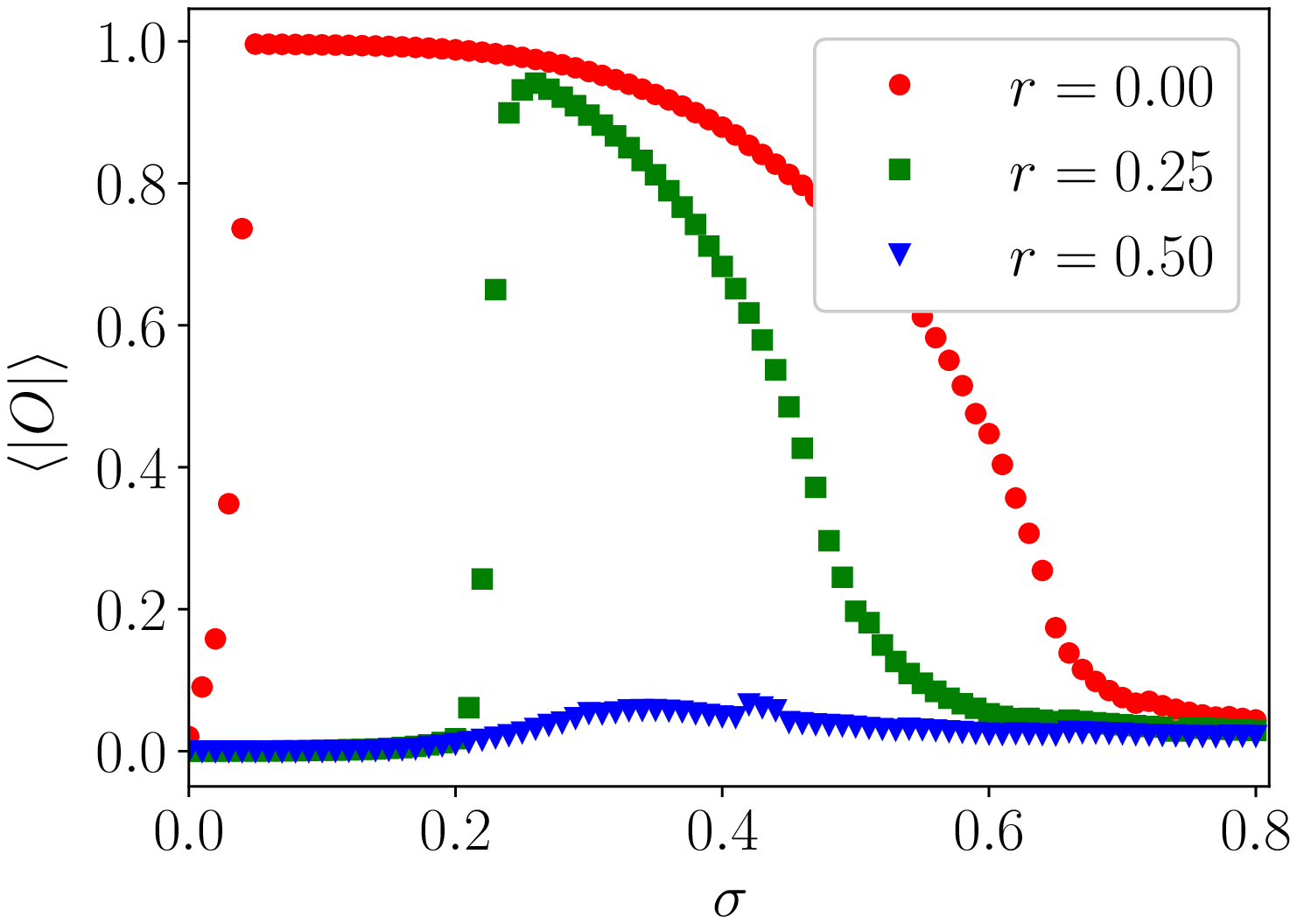}}
      \subfloat[BA]{\includegraphics[width=0.33\textwidth]{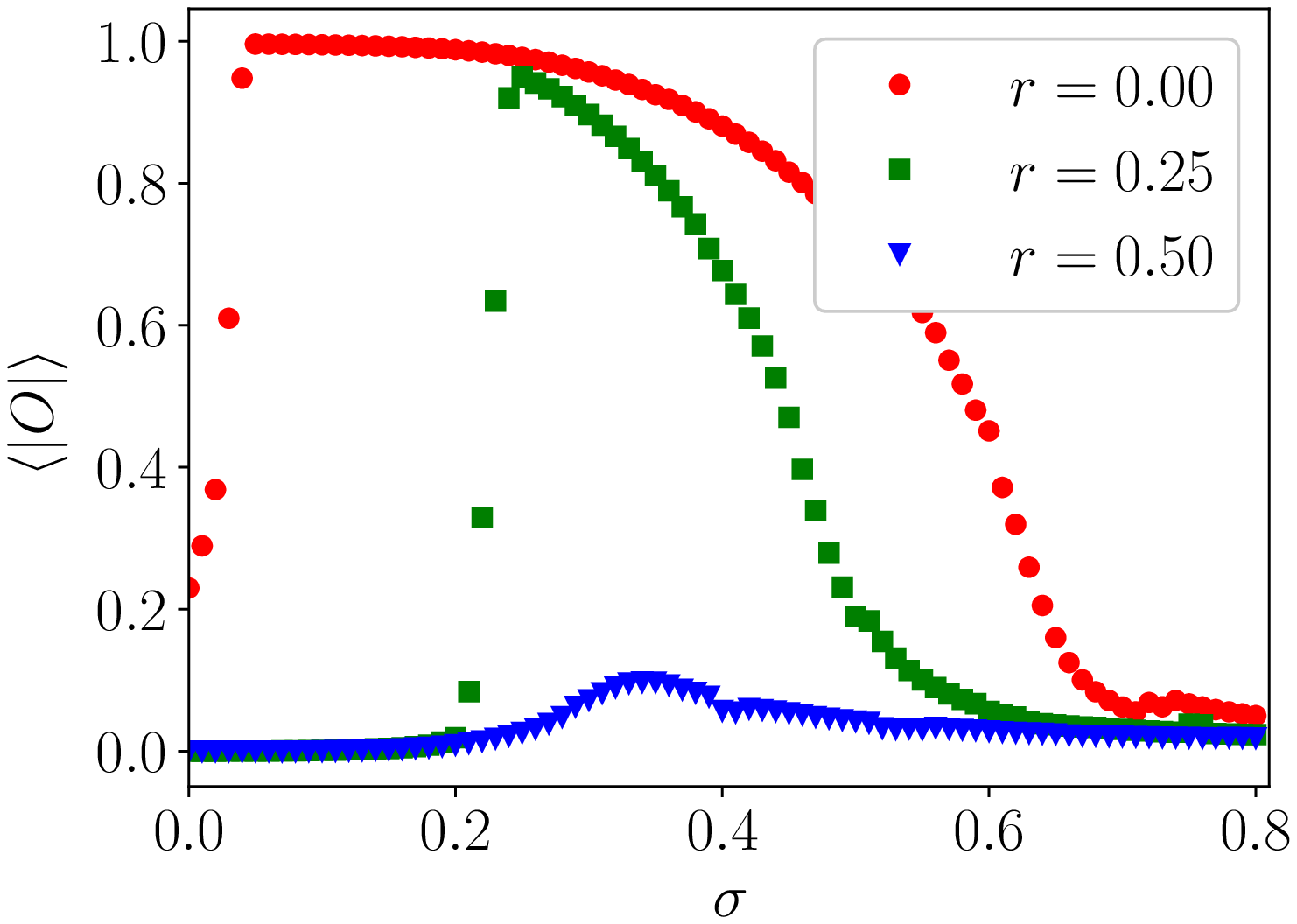}}
      \subfloat[WS]{\includegraphics[width=0.33\textwidth]{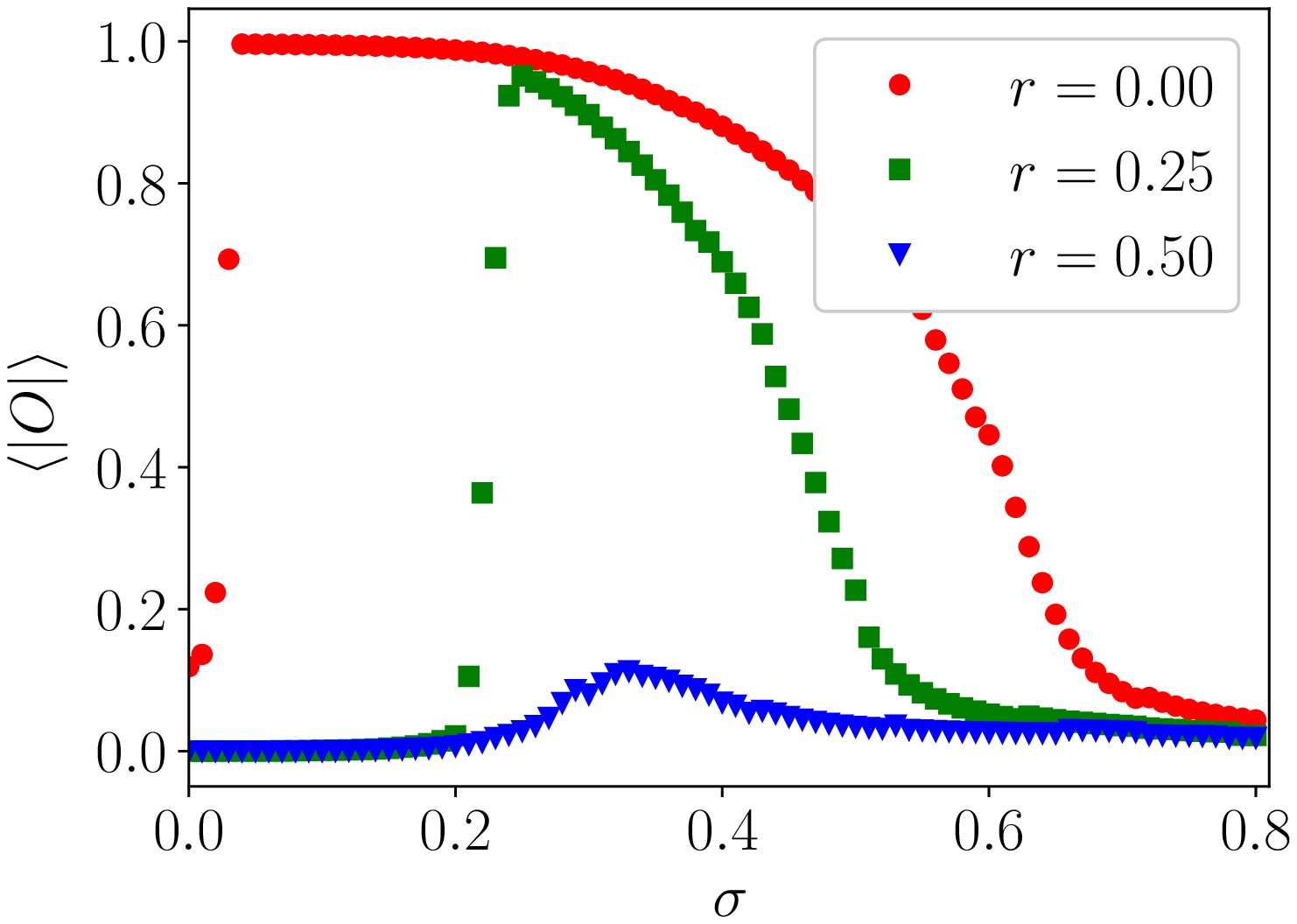}}
      \caption{Order induced by disorder for ER, WS and BA networks.   $\langle k \rangle = 50$, $\mu=0.34$ and different values of $r$. In the 1st disordered phase there is usually a coexistence of an ordered phase and a disordered phase. The only thing that changes is the fraction of samples that end-up ordered or disordered.}
      \label{fig:OxSig}
\end{figure*}

\begin{figure}[h]
      \centering
      \includegraphics[width=0.5\textwidth]{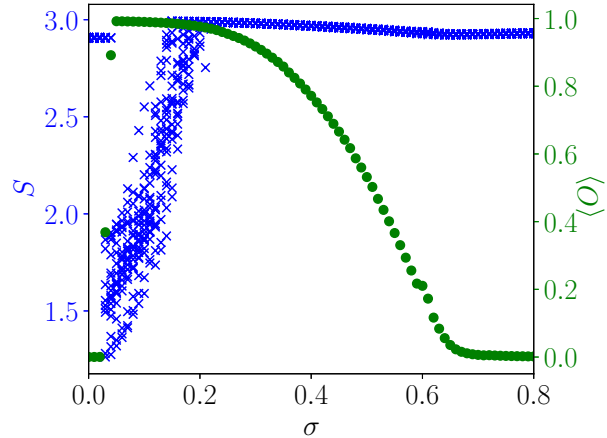}
      \caption{Entropy of the system, $S$, versus heterogeneity, $\sigma$ with the vertical axis to the left, for a Erd\H{o}s-R\'enyi network with $\mu = 0.34$ where the jump in the entropy is visible. The entropy was computed using the Kozachenko-Leonenko algorithm~\cite{1987kozachenkoL} and each $\times$ represents a sample. Average order parameter, $\langle O \rangle$, versus heterogeneity, $\sigma$, with the vertical axis to the right and $\bullet$ as marker.}
\label{fig:latent}
\end{figure}


Our Monte Carlo simulations are formulated in terms of a multi-agent system ~\cite{squazzoni2012agent,bianchi2015agent,klein2018agent} since individuals are the primary subject of social theories~\cite{conte2012manifesto}. The simulations here presented were performed considering $N = 10^4$ agents --- or nodes in the network --- and averaged over 50 samples. A new network was generated every 5 samples so that each data point is actually an average over 10 different networks generated using each model. All networks generated considered an average connectivity of $\langle k \rangle \approx 50$. By fixing the average connectivity we are able to better assess the relevance of the topology in the dynamics.

The dependence of the collective state $O$ on $\mu$ for both initial conditions --- $O(0) = 0$, $O(0) =1$ --- is visible in~\cref{fig:Oxsigma_l_1}.   For $\sigma=0$ --- which can be understood as the paradigmatic polarized case where the society is separated out into two perfectly aligned groups showing no dissension within them ---   there is a clear discontinuous transition between the reaching of a consensus and the impossibility thereof. We identify that class of transition taking into account the dependence of collective stationary state on the initial condition that leads to a hysteresis-like shape.   
The existence of hysteresis is a strong indication of the impact of the past (historical) social condition of the system --- which we can assign to the initial conditions of the present model --- and concurs with a phenomenon known in Sociology as social hysteresis~\cite{elster1976note}.    This aspect is emphasized by the increasing number of works showing its presence in many different social settings~\cite{2019oestereichPC,freitas2019imperfect,encinas2019majority,nowak2019homogeneous, encinas2018fundamental,chen2017first,gambaro2017influence,jkedrzejewski2017pair}. Importantly, none of these works showed the emergence of hysteresis under the dynamical interplay of multiple sources of disorder and interactions as done here. 4333

Besides hysteresis we have observed that in increasing the plurality of the groups by making $\sigma $ larger, it is possible to increase the value of the order parameter, i.e., booster the collective behavior (see~\cref{fig:OxSig}). That heterogeneity-assisted ordering shows a counterintuitive beneficial role in promoting different types of diversity (or let us say disorder) in a multi-agent system.   While reentrant transitions disorder-order-disorder have been reported in a mean-field zero-temperature Ginzburg-Landau under a quenched additive noise from a one-peak distribution~\cite{komin2010critical} --- i.e., without partisanship across the system --- the role of the disorder as a promoter of the collective stance is unknown for kinetic-like opinion dynamics. For instance, disorder manifested in the form of inflexibility in a fraction of the population only leads to a weakening of the collective stance~\cite{2015crokidakisO}.
Still in \cref{fig:OxSig}, we   see that increasing magnitudes of restrictions in the interactions, $r$, decreases the prominence of the ordered phase.

It is relevant to further explore the nature of the transition; as well-known in critical phenomena of equilibrium systems, the existence of a discontinuous phase transition is associated with a jump --- i.e., a discontinuity --- in the value of the entropy, $S$, of the system. In equilibrium statistical mechanics, that shift is   related to a latent heat $\Delta Q $ taking into account the relation $\Delta Q = T \, \Delta S$ where $T$ is the temperature of the system. We define that discontinuity in the entropy as latent heterogeneity in the system since it is unable to enhance the decrease in the cost of assuming the trivial collective stance, $O = 0$, over a nontrivial stance, $O \neq 1$. In \cref{fig:latent}, we depict the evolution of the entropy of the system $S \equiv - \int p(o) \, \ln p(o) \, do$ with respect to its heterogeneity, $\sigma $. Comparing to the diagram $O$ v $\sigma $ in, the circles with vertical axis to the right in the same figure, we verify that discontinuity.

\begin{figure}[t]
      \centering
      \captionsetup[subfigure]{justification=centering}
      \subfloat[$r=0$, ER, $O(0) = 0$]{\includegraphics[width=0.24\textwidth]{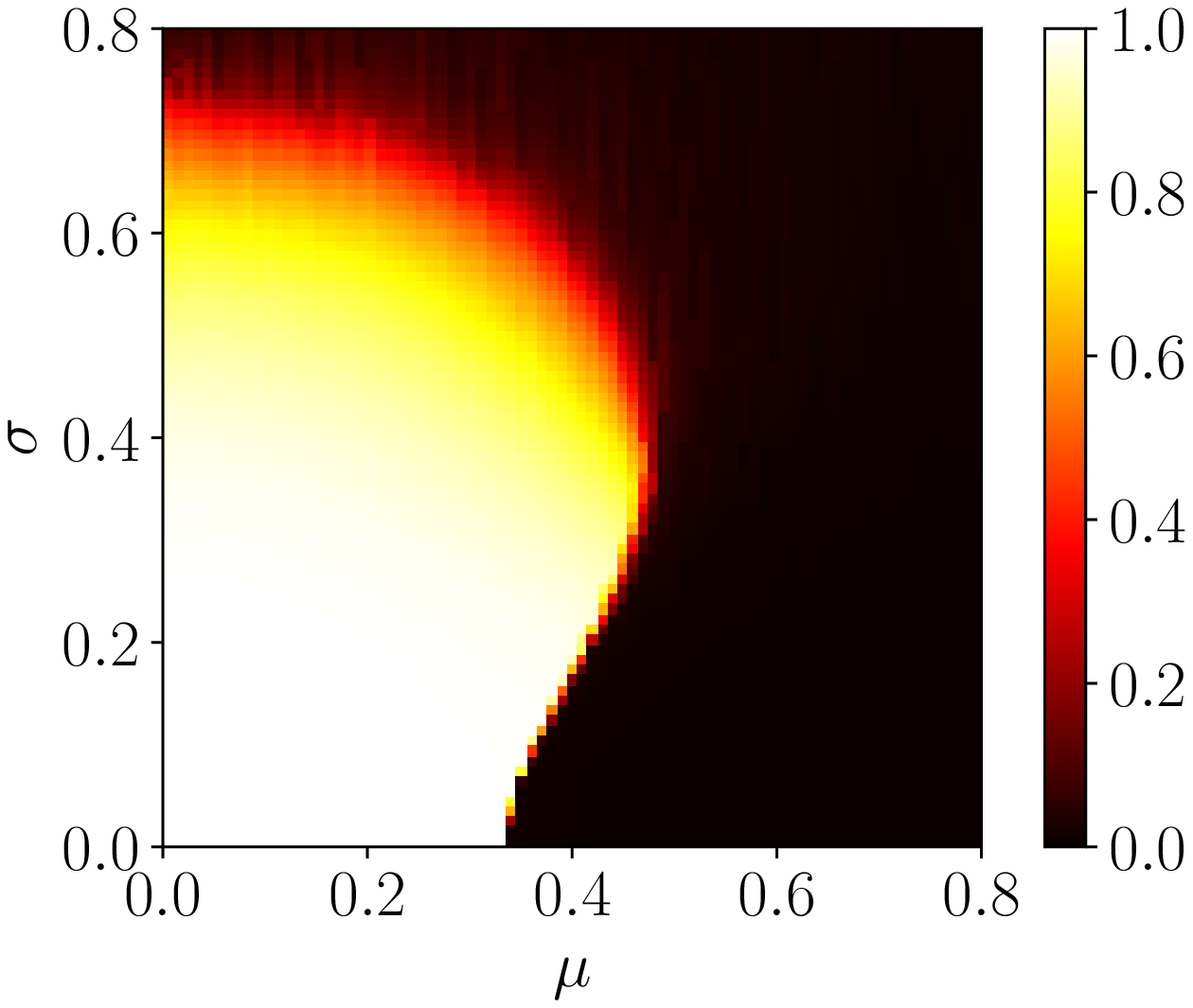}}
      \subfloat[$r=0$, ER, $O(0) = 1$]{\includegraphics[width=0.24\textwidth]{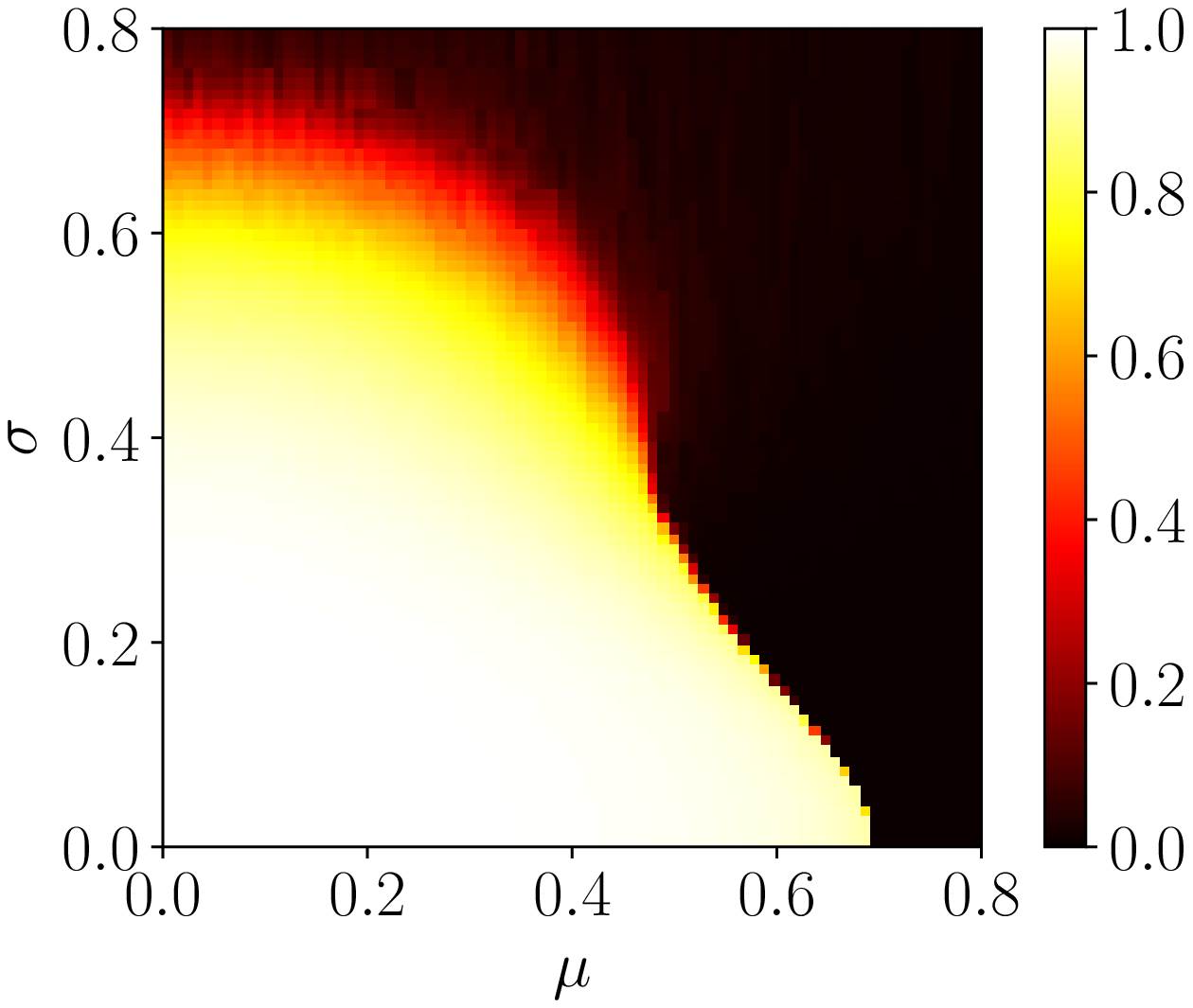}}
      \subfloat[$r=1/4,$ ER, $O(0) = 0$]{\includegraphics[width=0.24\textwidth]{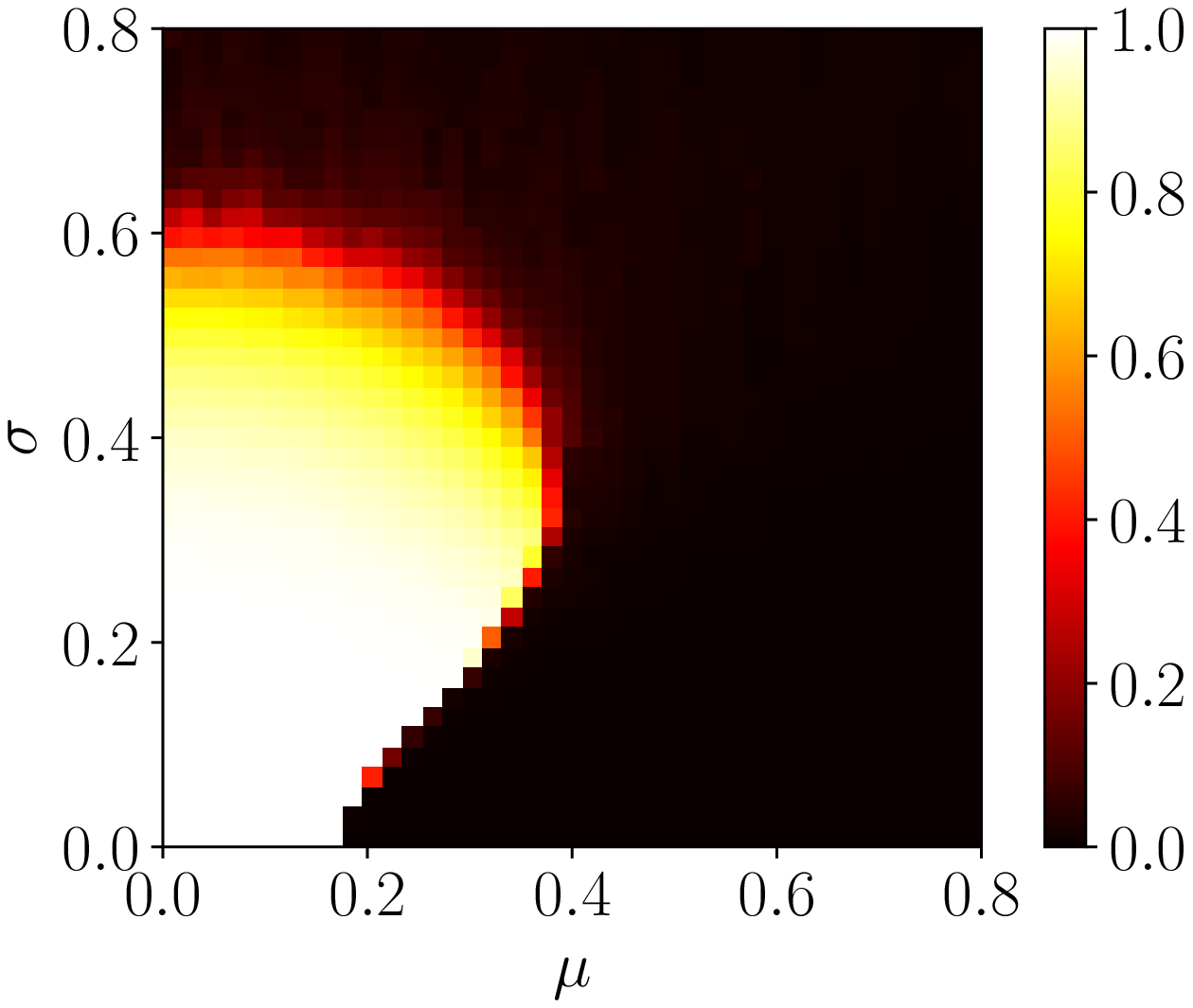}}
      \subfloat[$r=1/4,$ ER, $O(0) = 1$]{\includegraphics[width=0.24\textwidth]{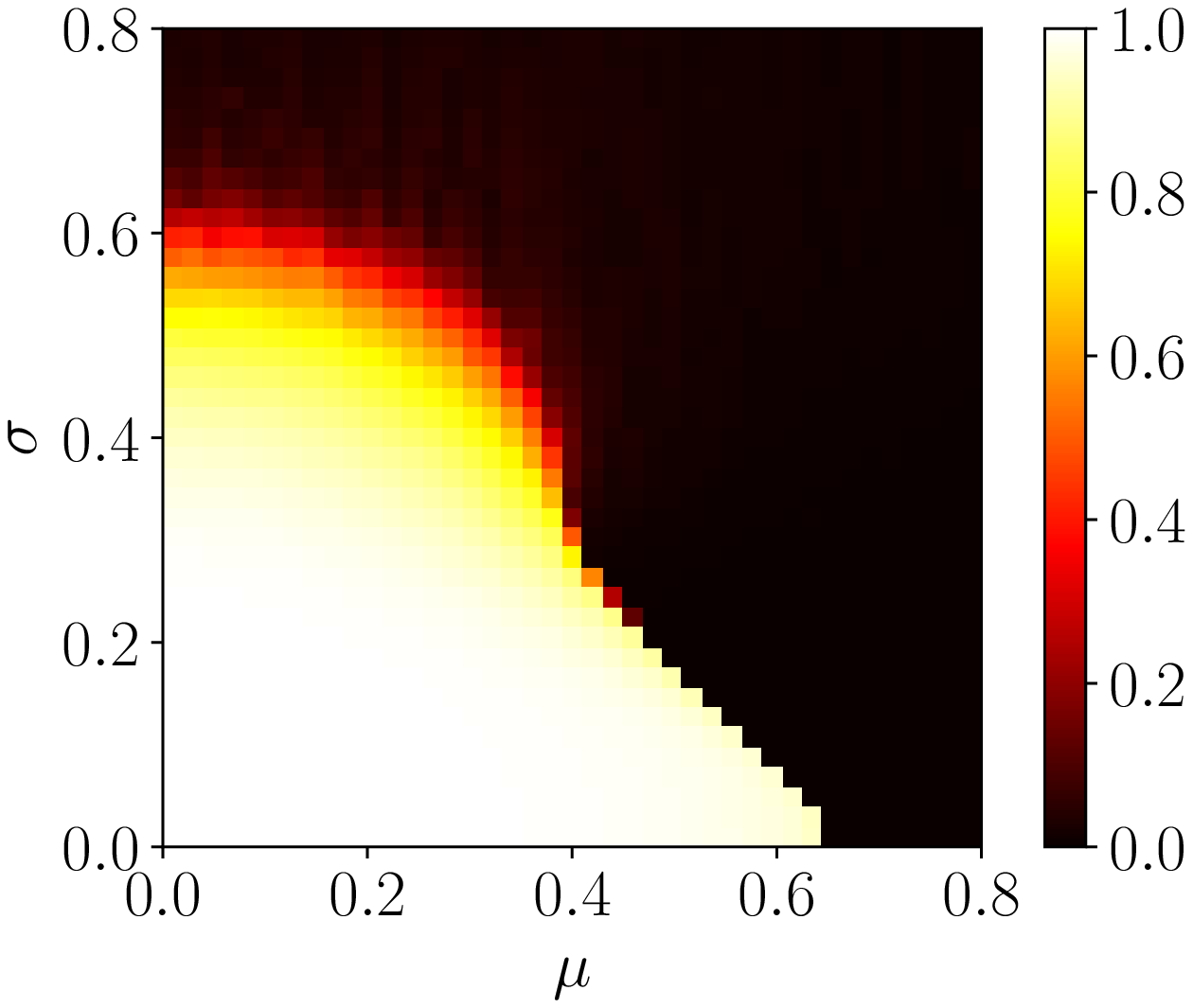}} \\
      \subfloat[$r=0$, BA, $O(0) = 0$]{\includegraphics[width=0.24\textwidth]{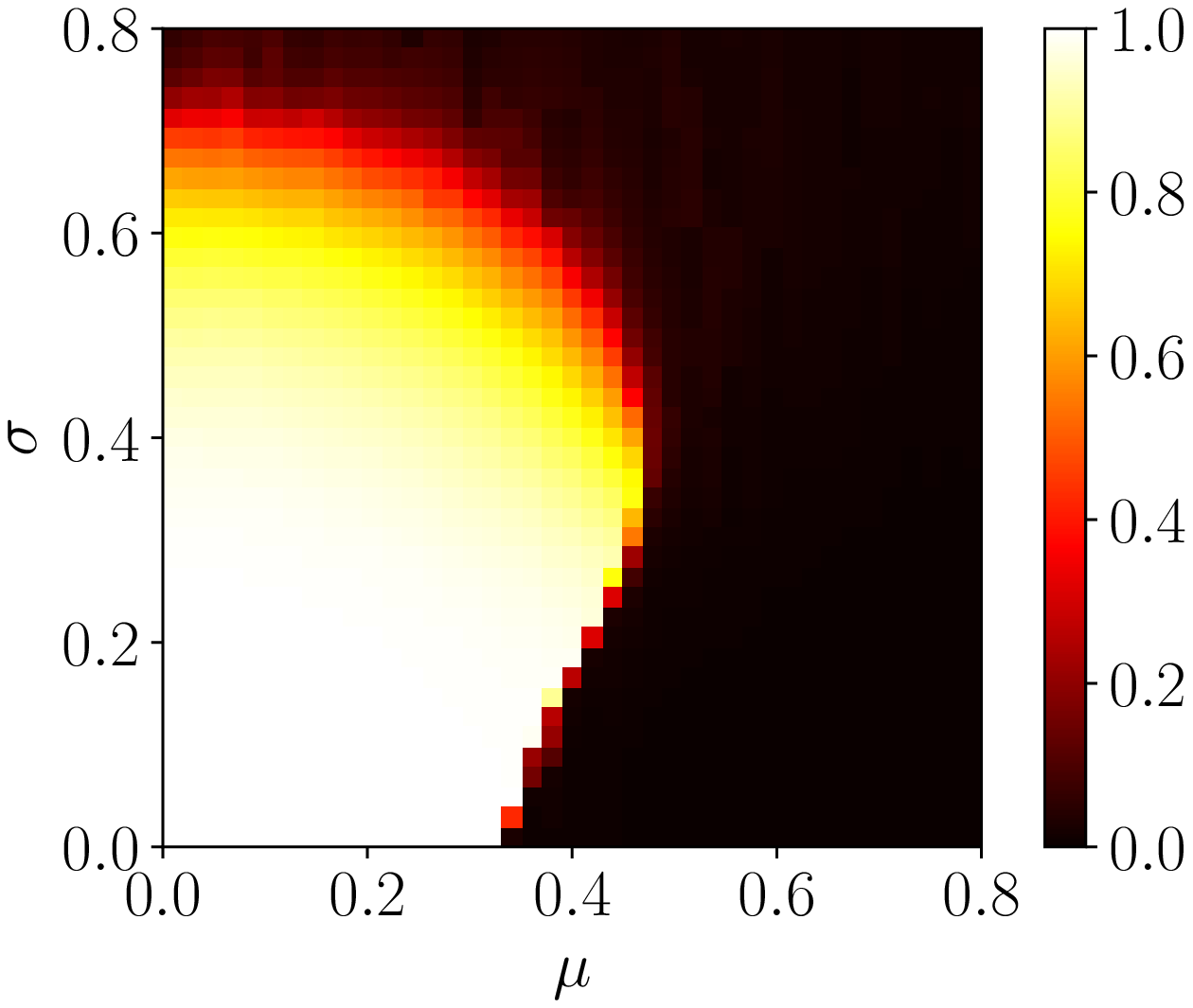}}
      \subfloat[$r=0$, BA, $O(0) = 1$]{\includegraphics[width=0.24\textwidth]{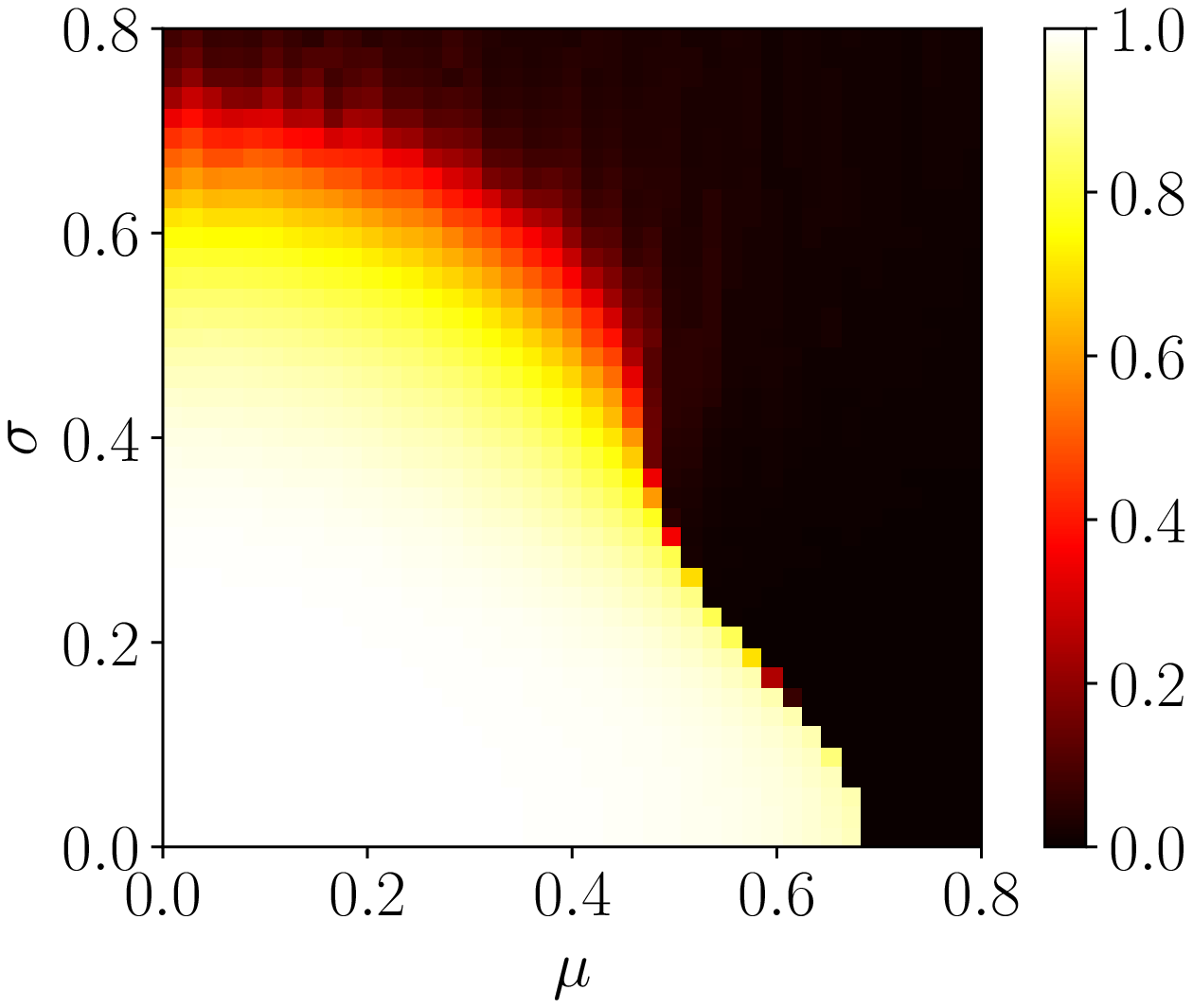}}
      \subfloat[$r=1/4,$ BA, $O(0) = 0$]{\includegraphics[width=0.24\textwidth]{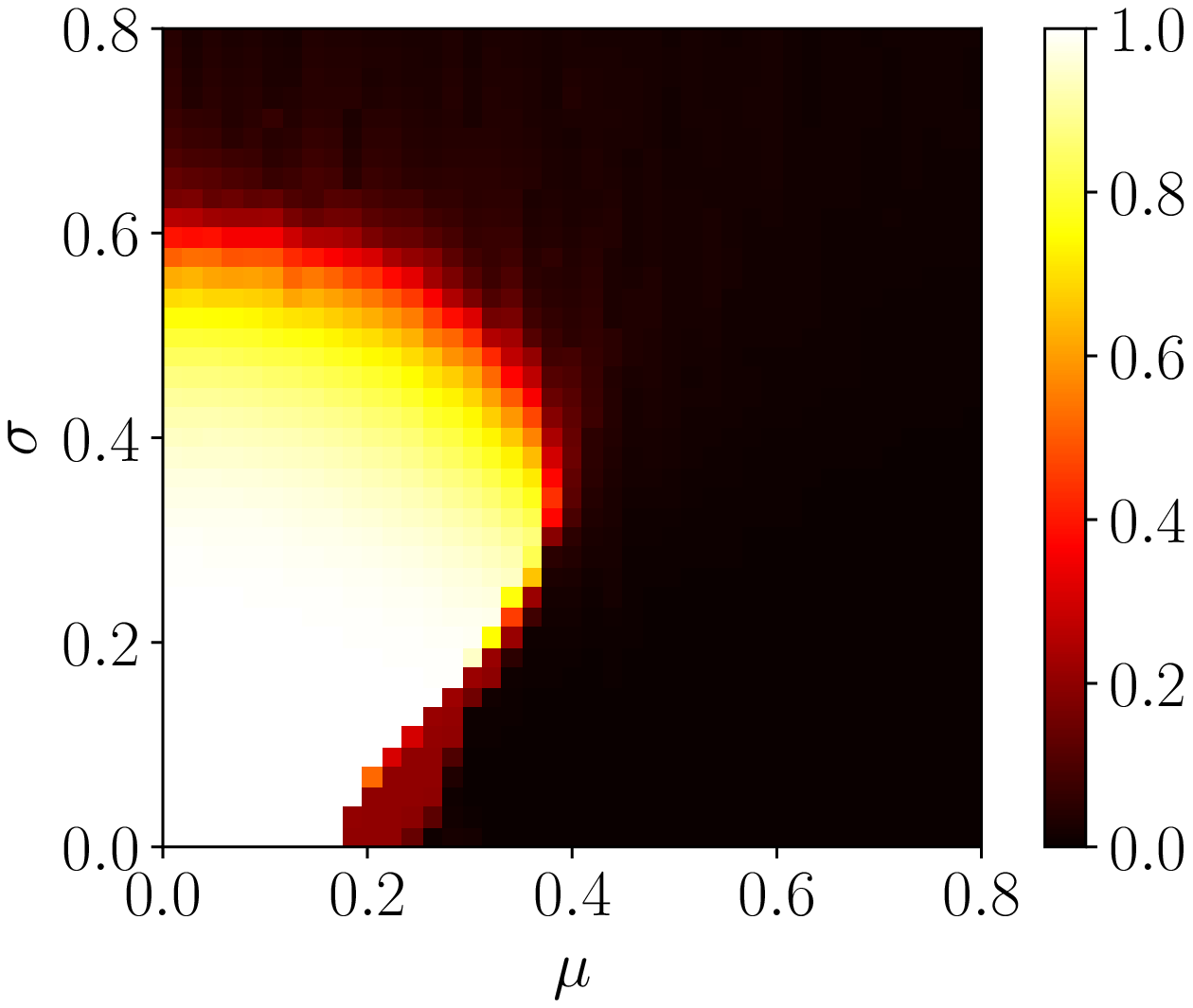}}
      \subfloat[$r=1/4,$ BA, $O(0) = 1$]{\includegraphics[width=0.24\textwidth]{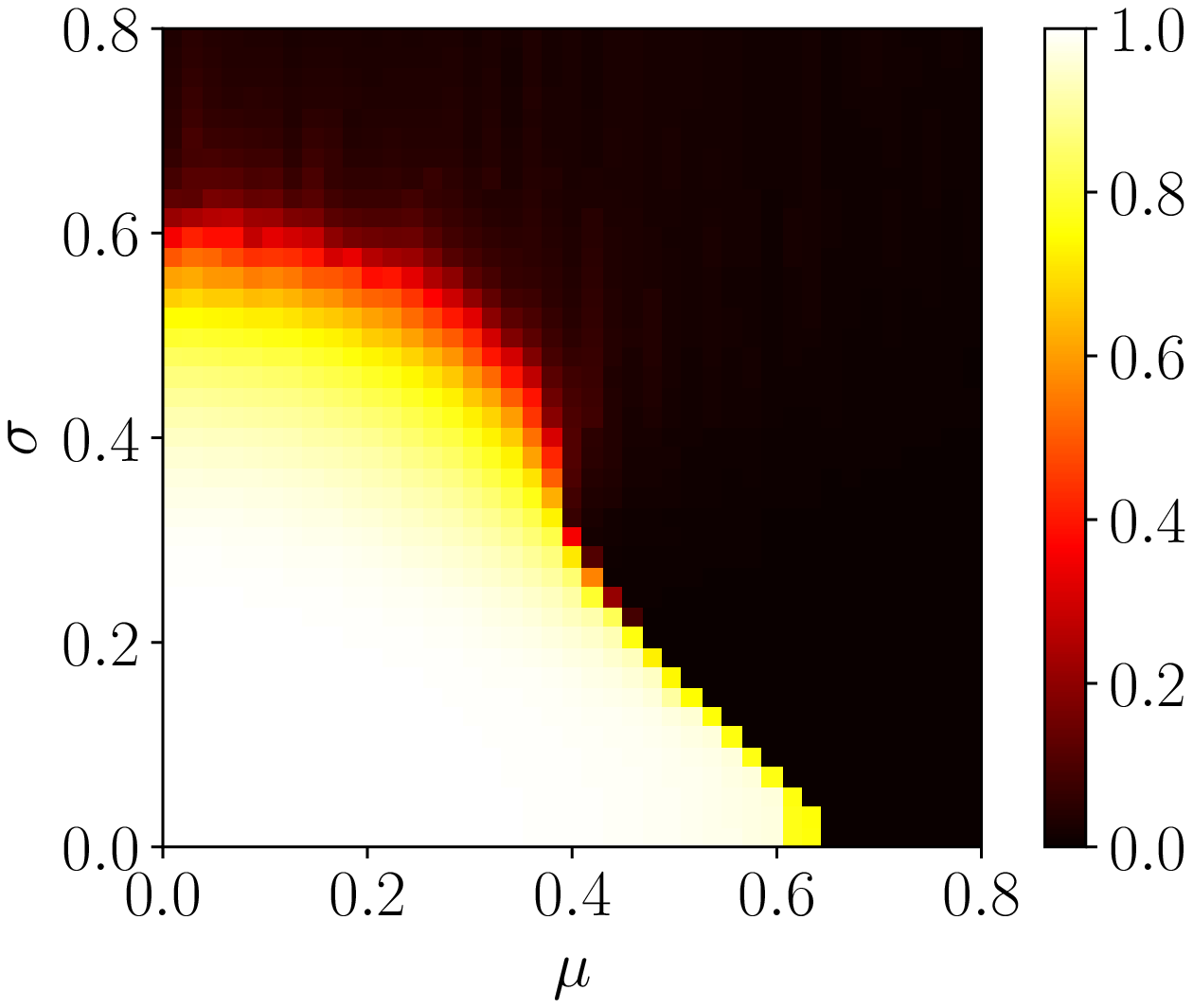}} \\
      \subfloat[$r=0$, WS, $O(0) = 0$]{\includegraphics[width=0.24\textwidth]{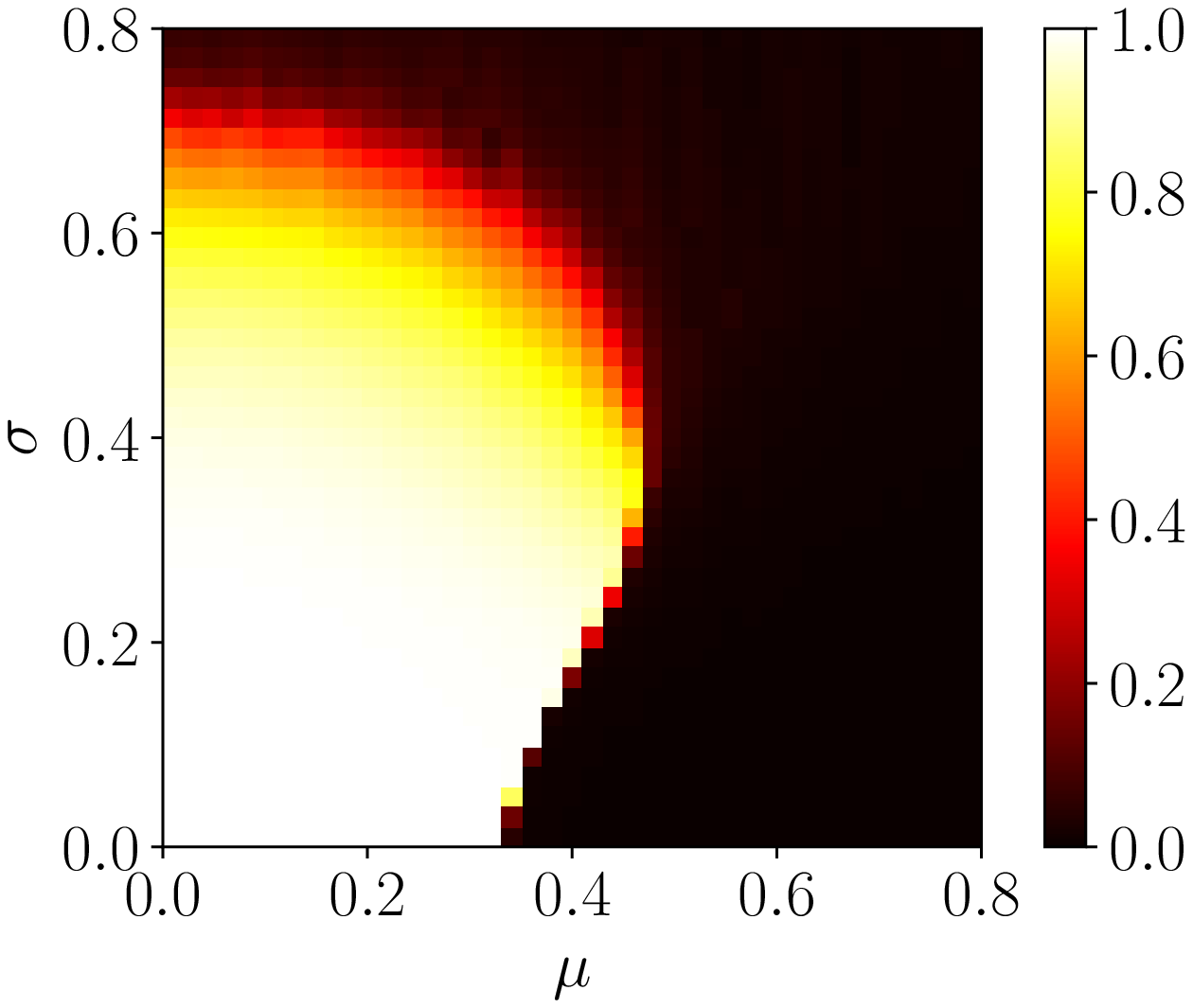}}
      \subfloat[$r=0$, WS, $O(0) = 1$]{\includegraphics[width=0.24\textwidth]{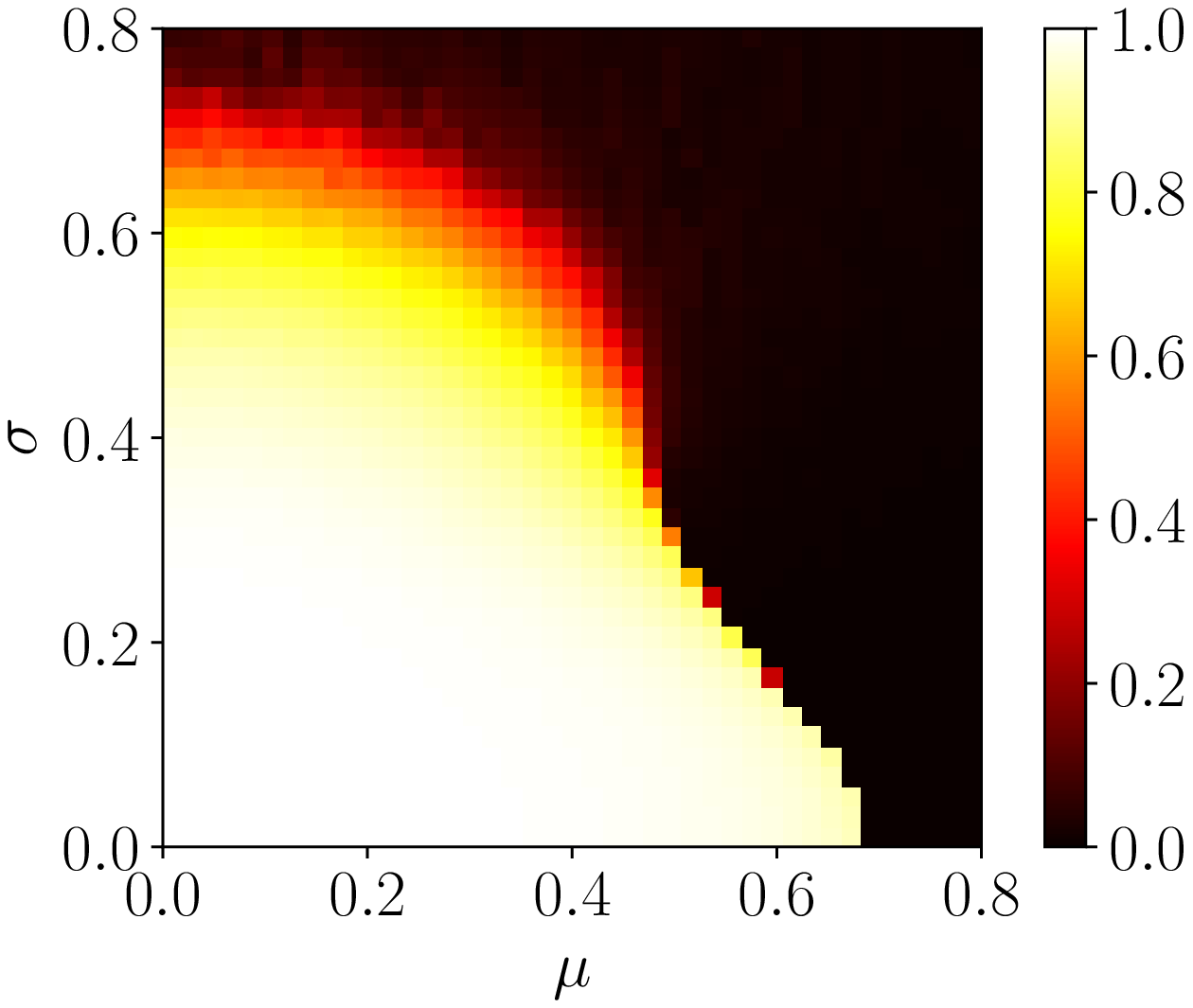}}                     
      \subfloat[$r=1/4,$ WS, $O(0) = 0$]{\includegraphics[width=0.24\textwidth]{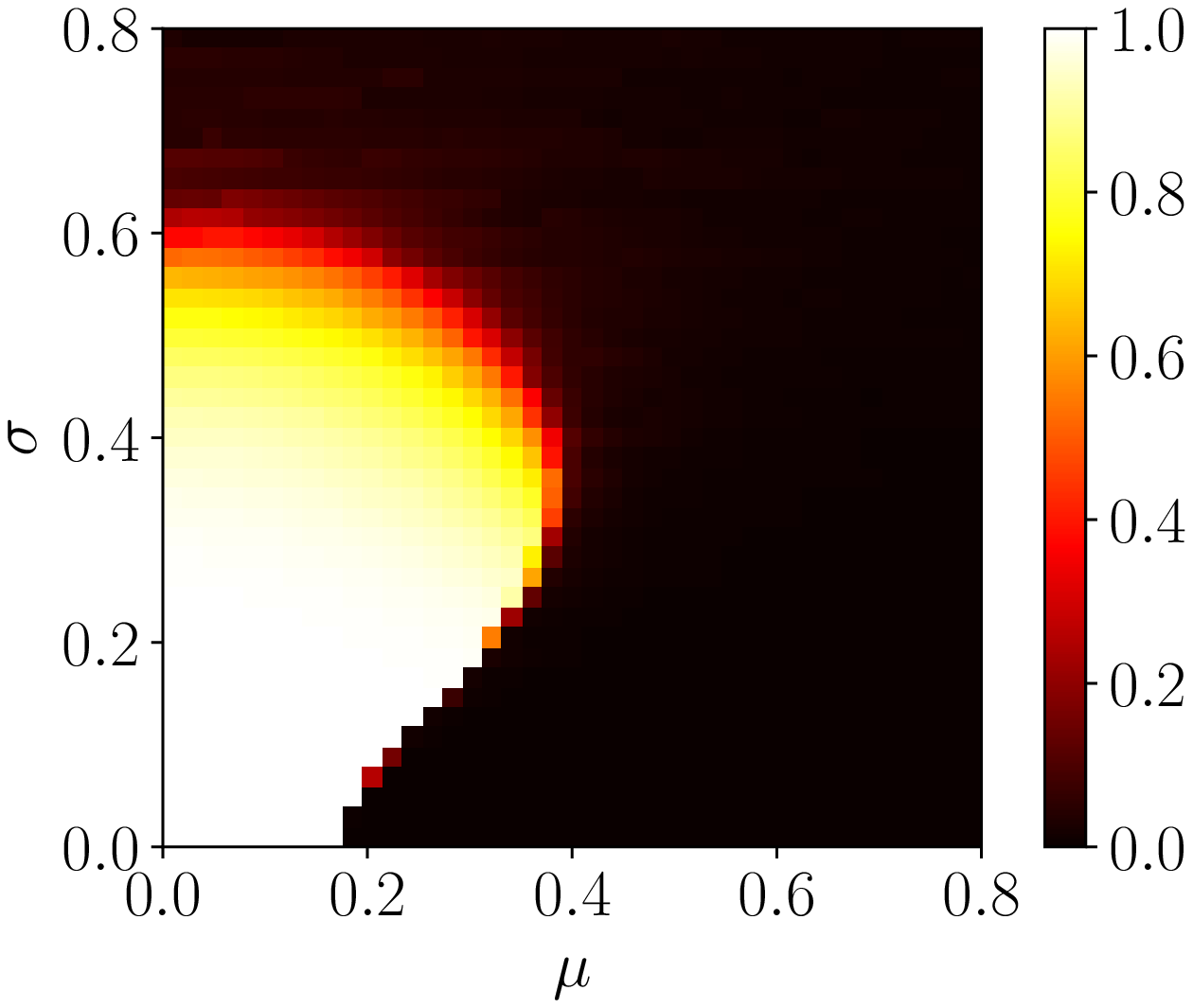}}
      \subfloat[$r=1/4,$ WS, $O(0) = 1$]{\includegraphics[width=0.24\textwidth]{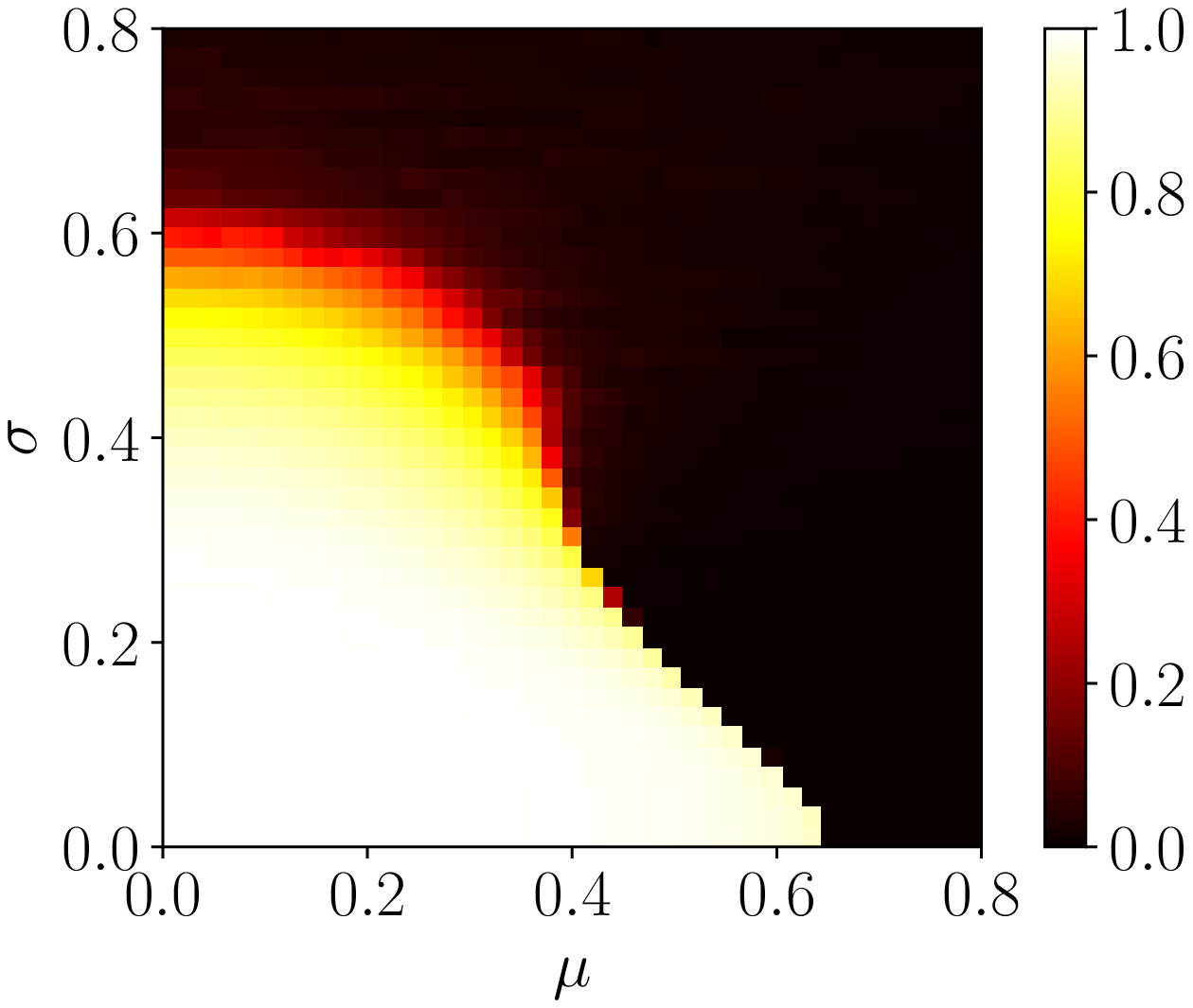}}      
      \caption{Order parameter diagram for diverse values of $\mu$ and $\sigma$ with $r=\{0,1/4\}$. Each row contains diagrams for a given type of network. Other parameters are specified in the subcaptions.}
      \label{fig:O_MuxSig_various}
\end{figure}


In \cref{fig:O_MuxSig_various} we address the robustness of our results. To perform this task we have carried out extensive Monte Carlo simulations considering $\mu$, $\sigma$, $r$ as free parameters and the two previously mentioned initial conditions. Each one of the diagrams has at least a total of 40x40x20 points of simulations. Within each row we show the diagrams for $O(0) = \{0,1\}$ and $r=\{0,1/4\}$. The main points of \cref{fig:O_MuxSig_various} can be summarized as:
\begin{itemize}
      \item The diagrams with initial condition $O(0)=0$ --- first and third columns --- show that the disorder-induced order is very robust. It appears across all network types, and for a range of both $\mu$ and $r$.
      %
      \item Each pair of initial conditions $O(0) = \{0,1\}$ presents a significant difference in the stationary state of the system for all networks and for both values of $r$. This confirms that the hysteresis is also robust to distinct $\mu$, $r$ and type of networks. This further underscores the indications of a discontinuous transition behavior. 
      \item Each pair $r=\{0,1/4\}$ (with a fixed $O(0)$ and type of network) confirms that increasing $r$ leads to a decrease in the area of the ordered phase. The origin of this inhibition of the collective stance due to $r$ can be traced back to the role of the decay rate $r$ in \cref{eq:evol,eq:smoothboundedconfidence}. It directly weakens the local term $\epsilon_1 g(t) o_j(t)$, which in turn, reduces the interactions between distant opinions and therefore the ability of the system to get to an ordered state.
\end{itemize}


Still in~\cref{fig:O_MuxSig_various}, but now focusing on the role of the structural disorder, take a look   at each triplet $\{ER,BA,WS\}$ of networks (with a fixed $O(0)$ and $r$). As aforementioned, for a fair comparison we keep the same average connectivity for those three types of networks. We see that the   breakdown of the uncorrelated features of the Erd\H{o}s-R\'enyi networks does not destroy neither the hysteresis nor the disorder-induced order. That is, such diagrams show that emergent phenomenology of our model is robust to correlations in the structural backbone of the network as can be seen in the diagrams for the BA and WS networks. Even though the overall features are qualitatively the same, there are some features to note: (i) the ER with its absence of clustering and degree-degree correlations leads to the largest area in the ordered state; (ii) the hub-and-spoke pattern in BA architectures leads to more ordered states than the presence of patterns manifested as triangulations and shortcuts across the WS network (that is decentralized).

About the order of the transition, the diagrams of~\cref{fig:O_MuxSig_various} exhibit regions undergoing an abrupt change in the order parameter. This result aligned with the hysteresis are indicators of a first-order phase transition, that in our case is of nonequilibrium nature. The diagrams also highlight that the increasing in $\sigma$ induces a smoothing in the phase transition. If the plurality parameter $\sigma$ increases too much, the transition changes from discontinuous to continuous with no signature of history-dependent effect. This is an agreement with Ref.~\cite{2016queiros}, in which it is considered   full-graph for modelling unbounded continuous opinions with individuals preference coming from a double gaussian.

Finally, a question deserves being addressed: what is the mechanism behind the phenomena we observed, namely the hysteresis and the ordered phase induced by disorder? Taking a final look at the \cref{eq:evol,eq:smoothboundedconfidence} we can   explain the origins of such phenomena by the imbalance in the   competition -- due to proper changes in the   combinations of the parameter space --   between two mechanisms: (i) the terms that tend to promote the global ordering such as the term associated to the local interaction $\epsilon_1 g(t) o_j(t)$ and the one associated to the global term  $\epsilon_2 O(t)$; (ii) the terms that tend to promote the permanence in the initial condition such as   $o_i(t)$ and the term associated to the   individual preference $h_i$.


\section{Final Remarks} \label{sec:remarks}

We have studied a dynamical model considering a continuous kinetic model in complex networks with restrictive interactions between agents,   modeled with a smooth bounded confidence. The agents are endued with a quenched individual preference~\cite{2016queiros} with plurality and polarization. Our model is able to reproduce a phenomenon previously found in field studies on Quantitative Sociology named social hysteresis. The main feature of our model is that it allows distinguishing different sorts of heterogeneity in the model so that in adjusting the impact of plurality over polarization --- which represents partisanship --- it is possible to control the type of transition between continuous and discontinuous transitions between a final non-trivial `{\it yea}/{\it nay}'   ($O \neq 0) $ and the trivial `{\it we agree to disagree}' ($O = 0$) collective stance. That behavior is robust across different network topologies with scale-free networks showing a larger volume of the space of parameters yielding $O \neq 0$.
Given the existence of hysteresis, we have asserted and confirmed that this behavior agrees to a jump in the entropy of the system that we defined as latent entropy; the term latent has to do with the incapacity of making the emergence of trivial over a nontrivial collective stance cheaper.

We have also observed the emergence of order induced by disorder, a typical phenomenon in magnetic systems. In addition, our results exhibit indicators of the occurrence of nonequilibrium first-order phase transitions, another uncommon behavior in models of opinion dynamics. Previous works showed the emergence of hysteresis, but to the best of our knowledge it is the first time that behavior is generated by the dynamical interplay of multiple sources of disorder and interactions as done here.

Within the field of opinion dynamics, our work has the contribution of presenting a novel mechanism for nonequilibrium hysteresis and for heterogeneity-assisted ordering. From a more practical perspective, opinion dynamics are important for other co-occurring processes such as vaccination dynamics where the success of a vaccination campaign also depends on the individual opinion about vaccines~\cite{2015wangAWW,2018piresOC}. In this sense, our work suggests that take into account the multifold interplay between several sources of disorder and interactions provides a better understanding of the collective stances which in turn can provide a better grasping of coupled opinion-vaccination dynamics.
In a future study we plan to pursue in details this   endeavour of research of coupling our present model of opinion dynamics -- with the insights we produce -- with a vaccination dynamics during an epidemic oubreak. The relevance of such coevolution spreading and others has been further stressed in a recent review~\cite{wang2019coevolution}. 


\section*{Acknowledgments}

The authors acknowledge financial support from the Brazilian funding agencies CNPq, CAPES and FAPERJ.


\section*{References}
\bibliography{hysteresis-opiniondynamics}

\begin{thebibliography}{10}
\expandafter\ifx\csname url\endcsname\relax
  \def\url#1{\texttt{#1}}\fi
\expandafter\ifx\csname urlprefix\endcsname\relax\def\urlprefix{URL }\fi
\expandafter\ifx\csname href\endcsname\relax
  \def\href#1#2{#2} \def\path#1{#1}\fi

\bibitem{barber2015causes}
M.~Barber, N.~McCarty, Causes and consequences of polarization, Political
  negotiation: A handbook 37 (2015) 39--43.
\newblock \href {https://doi.org/10.1017/CBO9781316091906.002}
  {\path{doi:10.1017/CBO9781316091906.002}}.

\bibitem{mccoy2018polarization}
J.~McCoy, T.~Rahman, M.~Somer, Polarization and the global crisis of democracy:
  Common patterns, dynamics, and pernicious consequences for democratic
  polities, American Behavioral Scientist 62~(1) (2018) 16--42.
\newblock \href {https://doi.org/10.1177/0002764218759576}
  {\path{doi:10.1177/0002764218759576}}.

\bibitem{urlpoliticalpolarization2020}
P.~R. Center,
  \href{https://www.people-press.org/2014/06/12/political-polarization-in-the-american-public}{Political
  Polarization in the American Public} (2014 (accessed: 10th February 2020)).
\newline\urlprefix\url{https://www.people-press.org/2014/06/12/political-polarization-in-the-american-public}

\bibitem{bottcher2020great}
L.~Böttcher, H.~Gersbach, The great divide: Drivers of polarization in the us
  public (2020).
\newblock \href {http://arxiv.org/abs/2001.05163} {\path{arXiv:2001.05163}}.

\bibitem{2018biswasLF}
S.~Biswas, F.~W.~S. Lima, O.~Flomenbom, Are socio-econo-physical models better
  to explain biases in societies?, Reports in Advances of Physical Sciences
  02~(02) (2018) 1850006.
\newblock \href {https://doi.org/10.1142/s2424942418500068}
  {\path{doi:10.1142/s2424942418500068}}.

\bibitem{sirbu2017opinion}
A.~S{\^\i}rbu, V.~Loreto, V.~D. Servedio, F.~Tria, Opinion dynamics: models,
  extensions and external effects, in: Participatory sensing, opinions and
  collective awareness, Springer, 2017, pp. 363--401.
\newblock \href {https://doi.org/10.1007/978-3-319-25658-0_17}
  {\path{doi:10.1007/978-3-319-25658-0_17}}.

\bibitem{sen2014sociophysics}
P.~Sen, B.~K. Chakrabarti, Sociophysics: an introduction, Oxford University
  Press, 2014.

\bibitem{2012galam}
S.~Galam, Sociophysics: a physicist's modeling of psycho-political phenomena,
  Springer Science \& Business Media, 2012.

\bibitem{2009castellanoFL}
C.~Castellano, S.~Fortunato, V.~Loreto, Statistical physics of social dynamics,
  Rev. Mod. Phys. 81 (2009) 591--646.
\newblock \href {https://doi.org/10.1103/RevModPhys.81.591}
  {\path{doi:10.1103/RevModPhys.81.591}}.

\bibitem{2008galam}
S.~Galam, Sociophysics: A review of galam models, International Journal of
  Modern Physics C 19~(03) (2008) 409--440.
\newblock \href {https://doi.org/10.1142/S0129183108012297}
  {\path{doi:10.1142/S0129183108012297}}.

\bibitem{2017albiPTZ}
G.~Albi, L.~Pareschi, G.~Toscani, M.~Zanella, Recent advances in opinion
  modeling: Control and social influence, in: Active Particles, Volume 1,
  Springer International Publishing, 2017, pp. 49--98.
\newblock \href {https://doi.org/10.1007/978-3-319-49996-3_2}
  {\path{doi:10.1007/978-3-319-49996-3_2}}.

\bibitem{2011xiaWX}
H.~Xia, H.~Wang, Z.~Xuan, Opinion dynamics, International Journal of Knowledge
  and Systems Science 2~(4) (2011) 72--91.
\newblock \href {https://doi.org/10.4018/jkss.2011100106}
  {\path{doi:10.4018/jkss.2011100106}}.

\bibitem{2013bellomoMT}
N.~Bellomo, G.~A. Marsan, A.~Tosin, Complex Systems and Society, Springer New
  York, 2013.
\newblock \href {https://doi.org/10.1007/978-1-4614-7242-1}
  {\path{doi:10.1007/978-1-4614-7242-1}}.

\bibitem{2002HegselmannK}
R.~Hegselmann, U.~Krause, Opinion dynamics and bounded confidence models,
  analysis, and simulation, Journal of Artificial Societies and Social
  Simulation 5~(3) (2002).

\bibitem{2010naldiPT}
G.~Naldi, L.~Pareschi, G.~Toscani (Eds.), Mathematical Modeling of Collective
  Behavior in Socio-Economic and Life Sciences, Birkh{\"a}user Boston, 2010.
\newblock \href {https://doi.org/10.1007/978-0-8176-4946-3}
  {\path{doi:10.1007/978-0-8176-4946-3}}.

\bibitem{dong2018survey}
Y.~Dong, M.~Zhan, G.~Kou, Z.~Ding, H.~Liang, A survey on the fusion process in
  opinion dynamics, Information Fusion 43 (2018) 57--65.
\newblock \href {https://doi.org/10.1016/j.inffus.2017.11.009}
  {\path{doi:10.1016/j.inffus.2017.11.009}}.

\bibitem{galam1990social}
S.~Galam, Social paradoxes of majority rule voting and renormalization group,
  Journal of Statistical Physics 61~(3-4) (1990) 943--951.
\newblock \href {https://doi.org/10.1007/BF01027314}
  {\path{doi:10.1007/BF01027314}}.

\bibitem{deffuant2000mixing}
G.~Deffuant, D.~Neau, F.~Amblard, G.~Weisbuch, Mixing beliefs among interacting
  agents, Advances in Complex Systems 3~(01n04) (2000) 87--98.
\newblock \href {https://doi.org/10.1142/S0219525900000078}
  {\path{doi:10.1142/S0219525900000078}}.

\bibitem{sen2012nonconservative}
Nonconservative kinetic exchange model of opinion dynamics with randomness and
  bounded confidence, Phys. Rev. E 86 (2012) 016115.
\newblock \href {https://doi.org/10.1103/PhysRevE.86.016115}
  {\path{doi:10.1103/PhysRevE.86.016115}}.

\bibitem{2018jedrzejewskiS}
A.~Jedrzejewski, K.~Sznajd-Weron, Impact of memory on opinion dynamics, Physica
  A: Statistical Mechanics and its Applications 505 (2018) 306 -- 315.
\newblock \href {https://doi.org/10.1016/j.physa.2018.03.077}
  {\path{doi:10.1016/j.physa.2018.03.077}}.

\bibitem{2009castellanoMP}
C.~Castellano, M.~A. Mu\~noz, R.~Pastor-Satorras, Nonlinear $q$-voter model,
  Phys. Rev. E 80 (2009) 041129.
\newblock \href {https://doi.org/10.1103/PhysRevE.80.041129}
  {\path{doi:10.1103/PhysRevE.80.041129}}.

\bibitem{anteneodo2017symmetry}
C.~Anteneodo, N.~Crokidakis, Symmetry breaking by heating in a continuous
  opinion model, Phys. Rev. E 95 (2017) 042308.
\newblock \href {https://doi.org/10.1103/PhysRevE.95.042308}
  {\path{doi:10.1103/PhysRevE.95.042308}}.

\bibitem{2019oestereichPC}
A.~L. Oestereich, M.~A. Pires, N.~Crokidakis, Three-state opinion dynamics in
  modular networks, Physical Review E 100~(3) (2019) 032312.
\newblock \href {https://doi.org/10.1103/physreve.100.032312}
  {\path{doi:10.1103/physreve.100.032312}}.

\bibitem{tessone2009diversity}
C.~J. Tessone, R.~Toral, Diversity-induced resonance in a model for opinion
  formation, The European Physical Journal B 71~(4) (2009) 549--555.
\newblock \href {https://doi.org/10.1140/epjb/e2009-00343-8}
  {\path{doi:10.1140/epjb/e2009-00343-8}}.

\bibitem{2016queiros}
S.~M.~D. Queir{\'{o}}s, Interplay between polarisation and plurality in a
  decision-making process with continuous opinions, Journal of Statistical
  Mechanics: Theory and Experiment 2016~(6) (2016) 063201.
\newblock \href {https://doi.org/10.1088/1742-5468/2016/06/063201}
  {\path{doi:10.1088/1742-5468/2016/06/063201}}.

\bibitem{2007jostGPO}
J.~T. Jost, J.~L. Napier, H.~Thorisdottir, S.~D. Gosling, T.~P. Palfai,
  B.~Ostafin, Are needs to manage uncertainty and threat associated with
  political conservatism or ideological extremity?, Personality and Social
  Psychology Bulletin 33~(7) (2007) 989--1007, pMID: 17620621.
\newblock \href {https://doi.org/10.1177/0146167207301028}
  {\path{doi:10.1177/0146167207301028}}.

\bibitem{2009jostFN}
J.~T. Jost, C.~M. Federico, J.~L. Napier, Political ideology: Its structure,
  functions, and elective affinities, Annual Review of Psychology 60~(1) (2009)
  307--337, pMID: 19035826.
\newblock \href {https://doi.org/10.1146/annurev.psych.60.110707.163600}
  {\path{doi:10.1146/annurev.psych.60.110707.163600}}.

\bibitem{2010hirshYXP}
J.~B. Hirsh, C.~G. DeYoung, X.~Xu, J.~B. Peterson, Compassionate liberals and
  polite conservatives: Associations of agreeableness with political ideology
  and moral values, Personality and Social Psychology Bulletin 36~(5) (2010)
  655--664, pMID: 20371797.
\newblock \href {https://doi.org/10.1177/0146167210366854}
  {\path{doi:10.1177/0146167210366854}}.

\bibitem{2014jostNAB}
J.~T. Jost, H.~H. Nam, D.~M. Amodio, J.~J. Van~Bavel, Political neuroscience:
  The beginning of a beautiful friendship, Political Psychology 35~(S1) (2014)
  3--42.
\newblock \href {https://doi.org/10.1111/pops.12162}
  {\path{doi:10.1111/pops.12162}}.

\bibitem{2010lallouacheCCC}
M.~Lallouache, A.~S. Chakrabarti, A.~Chakraborti, B.~K. Chakrabarti, Opinion
  formation in kinetic exchange models: Spontaneous symmetry-breaking
  transition, Phys. Rev. E 82 (2010) 056112.
\newblock \href {https://doi.org/10.1103/PhysRevE.82.056112}
  {\path{doi:10.1103/PhysRevE.82.056112}}.

\bibitem{2011biswasCC}
S.~Biswas, A.~K. Chandra, A.~Chatterjee, B.~K. Chakrabarti, Phase transitions
  and non-equilibrium relaxation in kinetic models of opinion formation, in:
  Journal of physics: conference series, Vol. 297, IOP Publishing, 2011, p.
  012004.
\newblock \href {https://doi.org/10.1088/1742-6596/297/1/012004}
  {\path{doi:10.1088/1742-6596/297/1/012004}}.

\bibitem{deffuant2004modelling}
G.~Deffuant, F.~Amblard, G.~Weisbuch, Modelling group opinion shift to extreme
  : the smooth bounded confidence model (2004).
\newblock \href {http://arxiv.org/abs/cond-mat/0410199}
  {\path{arXiv:cond-mat/0410199}}.

\bibitem{1997galam}
S.~Galam, Rational group decision making: A random field ising model at t = 0,
  Physica A: Statistical Mechanics and its Applications 238~(1) (1997) 66 --
  80.
\newblock \href {https://doi.org/10.1016/S0378-4371(96)00456-6}
  {\path{doi:10.1016/S0378-4371(96)00456-6}}.

\bibitem{1959erdosR}
P.~Erd\"{o}s, A.~R\'{e}nyi, On random graphs, i, Publicationes Mathematicae
  (Debrecen) 6 (1959) 290--297.

\bibitem{1999barabasiA}
A.-L. Barab{\'{a}}si, R.~Albert, Emergence of scaling in random networks,
  Science 286~(5439) (1999) 509--512.
\newblock \href {https://doi.org/10.1126/science.286.5439.509}
  {\path{doi:10.1126/science.286.5439.509}}.

\bibitem{1998wattsS}
D.~J. Watts, S.~H. Strogatz, Collective dynamics of `small-world' networks,
  Nature 393~(6684) (1998) 440--442.
\newblock \href {https://doi.org/10.1038/30918} {\path{doi:10.1038/30918}}.

\bibitem{stadtfeld2015micro}
C.~Stadtfeld, The micro--macro link in social networks, Emerging Trends in the
  Social and Behavioral Sciences: An Interdisciplinary, Searchable, and
  Linkable Resource (2015) 1--15\href
  {https://doi.org/10.1002/9781118900772.etrds0463}
  {\path{doi:10.1002/9781118900772.etrds0463}}.

\bibitem{1987kozachenkoL}
L.~Kozachenko, N.~N. Leonenko, Sample estimate of the entropy of a random
  vector, Problemy Peredachi Informatsii 23~(2) (1987) 9--16.

\bibitem{squazzoni2012agent}
F.~Squazzoni, Agent-based computational sociology, John Wiley \& Sons, 2012.
\newblock \href {https://doi.org/10.1002/9781119954200}
  {\path{doi:10.1002/9781119954200}}.

\bibitem{bianchi2015agent}
F.~Bianchi, F.~Squazzoni, Agent-based models in sociology, Wiley
  Interdisciplinary Reviews: Computational Statistics 7~(4) (2015) 284--306.
\newblock \href {https://doi.org/10.1002/wics.1356}
  {\path{doi:10.1002/wics.1356}}.

\bibitem{klein2018agent}
D.~Klein, J.~Marx, K.~Fischbach, Agent-based modeling in social science,
  history, and philosophy. an introduction, Historical Social
  Research/Historische Sozialforschung 43~(1 (163)) (2018) 7--27.
\newblock \href {https://doi.org/10.12759/hsr.43.2018.1.7-27}
  {\path{doi:10.12759/hsr.43.2018.1.7-27}}.

\bibitem{conte2012manifesto}
R.~Conte, N.~Gilbert, G.~Bonelli, C.~Cioffi-Revilla, G.~Deffuant, J.~Kertesz,
  V.~Loreto, S.~Moat, J.-P. Nadal, A.~Sanchez, et~al., Manifesto of
  computational social science, The European Physical Journal Special Topics
  214~(1) (2012) 325--346.
\newblock \href {https://doi.org/10.1140/epjst/e2012-01697-8}
  {\path{doi:10.1140/epjst/e2012-01697-8}}.

\bibitem{elster1976note}
J.~Elster, A note on hysteresis in the social sciences, Synthese 33~(1) (1976)
  371--391.
\newblock \href {https://doi.org/10.1007/BF00485452}
  {\path{doi:10.1007/BF00485452}}.

\bibitem{freitas2019imperfect}
F.~Freitas, A.~R. Vieira, C.~Anteneodo, Imperfect bifurcations in opinion
  dynamics under external fields, Journal of Statistical Mechanics: Theory and
  Experiment (2020).
\newblock \href {https://doi.org/10.1088/1742-5468/ab6848}
  {\path{doi:10.1088/1742-5468/ab6848}}.

\bibitem{encinas2019majority}
J.~Encinas, H.~Chen, M.~M. de~Oliveira, C.~E. Fiore, Majority vote model with
  ancillary noise in complex networks, Physica A: Statistical Mechanics and its
  Applications 516 (2019) 563--570.
\newblock \href {https://doi.org/10.1016/j.physa.2018.10.055}
  {\path{doi:10.1016/j.physa.2018.10.055}}.

\bibitem{nowak2019homogeneous}
B.~Nowak, K.~Sznajd-Weron, Homogeneous symmetrical threshold model with
  nonconformity: Independence versus anticonformity, Complexity 2019 (2019).
\newblock \href {https://doi.org/10.1155/2019/5150825}
  {\path{doi:10.1155/2019/5150825}}.

\bibitem{encinas2018fundamental}
J.~M. Encinas, P.~E. Harunari, M.~de~Oliveira, C.~E. Fiore, Fundamental
  ingredients for discontinuous phase transitions in the inertial majority vote
  model, Scientific reports 8~(1) (2018) 9338.
\newblock \href {https://doi.org/10.1038/s41598-018-27240-4}
  {\path{doi:10.1038/s41598-018-27240-4}}.

\bibitem{chen2017first}
H.~Chen, C.~Shen, H.~Zhang, G.~Li, Z.~Hou, J.~Kurths, First-order phase
  transition in a majority-vote model with inertia, Phys. Rev. E 95 (2017)
  042304.
\newblock \href {https://doi.org/10.1103/PhysRevE.95.042304}
  {\path{doi:10.1103/PhysRevE.95.042304}}.

\bibitem{gambaro2017influence}
J.~P. Gambaro, N.~Crokidakis, The influence of contrarians in the dynamics of
  opinion formation, Physica A: Statistical Mechanics and its Applications 486
  (2017) 465--472.
\newblock \href {https://doi.org/10.1016/j.physa.2017.05.040}
  {\path{doi:10.1016/j.physa.2017.05.040}}.

\bibitem{jkedrzejewski2017pair}
A.~Jedrzejewski, Pair approximation for the q-voter model with independence on
  complex networks, Phys. Rev. E 95 (2017) 012307.
\newblock \href {https://doi.org/10.1103/PhysRevE.95.012307}
  {\path{doi:10.1103/PhysRevE.95.012307}}.

\bibitem{komin2010critical}
N.~Komin, L.~Lacasa, R.~Toral, Critical behavior of a ginzburg--landau model
  with additive quenched noise, Journal of Statistical Mechanics: Theory and
  Experiment 2010~(12) (2010) P12008.
\newblock \href {https://doi.org/10.1088/1742-5468/2010/12/P12008}
  {\path{doi:10.1088/1742-5468/2010/12/P12008}}.

\bibitem{2015crokidakisO}
N.~Crokidakis, P.~M.~C. de~Oliveira, Inflexibility and independence: Phase
  transitions in the majority-rule model, Physical Review E 92~(6) (2015)
  062122.
\newblock \href {https://doi.org/10.1103/physreve.92.062122}
  {\path{doi:10.1103/physreve.92.062122}}.

\bibitem{2015wangAWW}
Z.~Wang, M.~A. Andrews, Z.-X. Wu, L.~Wang, C.~T. Bauch, Coupled
  disease{\textendash}behavior dynamics on complex networks: A review, Physics
  of Life Reviews 15 (2015) 1--29.
\newblock \href {https://doi.org/10.1016/j.plrev.2015.07.006}
  {\path{doi:10.1016/j.plrev.2015.07.006}}.

\bibitem{2018piresOC}
M.~A. Pires, A.~L. Oestereich, N.~Crokidakis, Sudden transitions in coupled
  opinion and epidemic dynamics with vaccination, Journal of Statistical
  Mechanics: Theory and Experiment 2018~(5) (2018) 053407.
\newblock \href {https://doi.org/10.1088/1742-5468/aabfc6}
  {\path{doi:10.1088/1742-5468/aabfc6}}.

\bibitem{wang2019coevolution}
W.~Wang, Q.-H. Liu, J.~Liang, Y.~Hu, T.~Zhou, Coevolution spreading in complex
  networks, Physics Reports (2019).
\newblock \href {https://doi.org/10.1016/j.physrep.2019.07.001}
  {\path{doi:10.1016/j.physrep.2019.07.001}}.

\end{thebibliography}

\end{document}